%% file: LU2012xxxa.tex
\documentclass[
aps,%
12pt,%
final,%
notitlepage,%
oneside,%
onecolumn,%
nobibnotes,%
nofootinbib,%
superscriptaddress,%
noshowpacs,%
centertags]%
{revtex4}

\usepackage{lscape}
\usepackage{multirow}
\usepackage{longtable}
\usepackage{epsfig}

\begin{document}

\pagenumbering{arabic}

\title{Atlas and wavenumber tables for visible part of the rovibronic multiline 
emission spectrum of the $D_2$ molecule. I. Wavenumber range $23894 \div 18161$ cm$^{-1}$ 
measured with moderate resolution.}

\author{B.~P.~Lavrov}
\email{lavrov@pobox.spbu.ru}
\author{I.~S.~Umrikhin}
\affiliation{
Faculty of Physics, St.-Petersburg State University, \\
St.-Petersburg, 198504, Russia}

\begin{abstract}
The visible part ($\approx 419 \div 550$ nm) of the 
multiline electronic-vibro-rotational emission spectrum of the $D_2$ molecule 
was recorded with moderate resolution (line widths $\approx 0.013$ nm). 
The resolution was limited by Doppler broadening of 
spectral lines. After numerical deconvolution of the recorded intensity 
distributions and proper calibration of the spectrometer the new set of 
wavenumber values was obtained. The results are reported in the form of an 
atlas divided into 36 sections covering about 1.5 nm, containing pictures of 
images in the focal plane of the spectrometer, intensity distributions in 
linear and logarithmic scales and the table containing wavenumber and relative 
intensity values for 6545 spectral lines together with existing line assignments.
\end{abstract}

\maketitle

\section*{Introduction}

Any activity in practical spectroscopy starts from recording certain spectra 
and recognizing lines, branches and bands interesting for an experimentalist. 
Most straightforward, dependable and easy way for the recognition is a comparison 
of an observed spectrum with certain reference atlases of spectra for various 
atoms and molecules. Currently for molecular deuterium such atlas is available 
only for limited part of vacuum ultraviolet (VUV) emission spectrum 
$78.60 \div 171.35$ nm. Present work reports atlas of multiline 
electronic-vibro-rotational (rovibronic) spectrum of the $D_2$ molecule for visible 
part ($419 \div 550$ nm) of the emission spectrum most suitable for practical 
applications in studies of deuterium containing plasmas. 

Experimental studies of the $D_2$ spectrum have been started soon after discovery of 
the heavy isotope of atomic hydrogen \cite{Urey1932_1, Urey1932_2}. Wavenumber values 
of rovibronic radiative transitions obtained by emission 
spectroscopy in visible 
\cite{DiekeBlue1935, Dieke1935_1, Dieke1935_2, Dieke1935_3, Dieke1936, DiekeLewis1937}
and infrared (IR) \cite{DiekePorto, GloersenDieke, DiekeCunningham} parts of the $D_2$ 
spectrum together with those obtained by VUV 
\cite{Wilkinson1968, Monfils1968, BredohlHerzberg1972, DabrowskiHerzberg1974, TT1975, LLR1980}
and anticrossing 
\cite{JLDRFMZ1976, MFZ1976, MZF1978, MF1977, FMZ1976} spectroscopic experiments were 
collected and analyzed in the review paper \cite{FSC1985}. Later fragmentary measurements were made 
in middle infrared (about $4.5$ $\mu m$) by FTIR (Fourier transform infrared) 
\cite{DabrHerz} and laser \cite{Davies} spectroscopy. Measurements of the wavenumber 
values for separate rovibronic lines and empirical determination of singlet rovibronic 
term values are in progress up to now 
\cite{RLTBjcp2006, RLTBjcp2007, RIVLTUmol2008, LSJUMchp2010, GJRT2011, DIUROJNTGSKEmol2011}. 

\begin{figure}[!ht]
\begin{center}
\includegraphics[width=\textwidth]{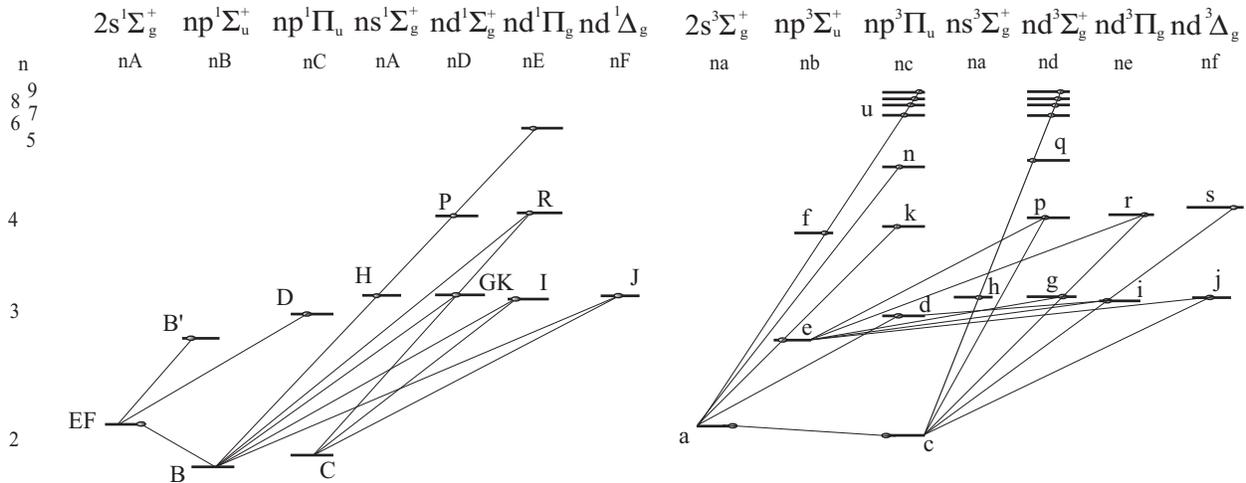}
\caption{
Grotrian diagram of currently known 
electronic states and studied band systems of the $D_2$ molecule
according to \cite{FSC1985}.}\label{allgrotr}
\end{center}
\end{figure}

The spectrum of the $D_2$ molecule is caused by both singlet-singlet and triplet-triplet rovibronic 
transitions. The intercombination lines were not observed yet. The most interesting 
resonance singlet-to-singlet band systems connected with ground electronic state are 
located in vacuum ultraviolet. Singlet-to-singlet and triplet-to-triplet transitions 
between excited electronic states are responsible for light emission of ionized gases and 
plasmas in near infrared, visible and near ultraviolet. Grotrian diagram of currently known 
electronic states and studied band systems of the $D_2$ molecule is shown in Fig.~\ref{allgrotr}.
It should be noted that 
visible part of the spectrum is most often used for spectroscopic diagnostics 
of non-equilibrium plasmas \cite{GLT1982, PBS2005, LKOR1997, LMKR1999, RDKL2001, LPR2006}. 

{\def\baselinestretch{1.05}
\setlength{\tabcolsep}{3pt}

\begin{longtable}[]{ccrlcrl}
\caption[]{The list of notations used for designation of electronic states of the $D_2$ 
molecule corresponding to various electron configurations.}\label{notations}
\endfirsthead
\multicolumn{6}{l}{Table \ref{notations} (Continued).}\\ 

\hline
Electron & 
\multicolumn{3}{c}{Singlet electronic states} & 
\multicolumn{3}{c}{Triplet electronic states} \\
\cline{2-7}
configuration  & 
   \multicolumn{2}{c}{Traditional} & Dieke &
   \multicolumn{2}{c}{Traditional} & Dieke \\
\cline{2-7}
 & \cite{Huber} & \multicolumn{2}{c}{\cite{FSC1985}} &
   \cite{Huber} & \multicolumn{2}{c}{\cite{FSC1985}} \\

\hline
\endhead

\hline
Electron & 
\multicolumn{3}{c}{Singlet electronic states} & 
\multicolumn{3}{c}{Triplet electronic states} \\
\cline{2-7}
configuration  & 
   \multicolumn{2}{c}{Traditional} & Dieke &
   \multicolumn{2}{c}{Traditional} & Dieke \\
\cline{2-7}
 & \cite{Huber} & \multicolumn{2}{c}{\cite{FSC1985}} &
   \cite{Huber} & \multicolumn{2}{c}{\cite{FSC1985}} \\
\hline

$1s\sigma^2$         & $X^1\Sigma_g^+ 1s\sigma$   & $X(1s)^1\Sigma_g^+$    & $1A$     & \multicolumn{3}{c}{$-$} \\
                                                                       
$1s\sigma 2s\sigma$  & $E^1\Sigma_g^+ 2s\sigma$   & 
                                    \multirow{2}{*}{$EF^1\Sigma_g^+$}      & 
                                                             \multirow{2}{*}{$EF$}    & $a^3\Sigma_g^+ 2s\sigma$   & $a(2s)^3\Sigma_g^+$   & $2a$     \\
$2p\sigma^2$         & $F^1\Sigma_g^+ 2p\sigma^2$ &                        &          &                            &                       &          \\
$1s\sigma 2p\sigma$  & $B^1\Sigma_u^+ 2p\sigma$   & $B(2p)^1\Sigma_u^+$    & $2B$     & $b^3\Sigma_u^+ 2p\sigma$   & $b(2p)^3\Sigma_u^+$   & $2b$     \\
$1s\sigma 2p\pi$     & $C^1\Pi_u 2p\pi$           & $C(2p)^1\Pi_u^\pm$     & $2C^\pm$ & $c^3\Pi_u 2p\pi$           & $c(2p)^3\Pi_u^\pm$    & $2c^\pm$ \\
                                                                       
$1s\sigma 3s\sigma$  &                            & $H(3s)^1\Sigma_g^+$    & $3A$     &                            & $h(3s)^3\Sigma_g^+$   & $3a$     \\
$1s\sigma 3p\sigma$  & $B'^1\Sigma_u^+ 3p\sigma$  & $B'(3p)^1\Sigma_u^+$   & $3B$     & $e^3\Sigma_u^+ 3p\sigma$   & $e(3p)^3\Sigma_u^+$   & $3b$     \\
$1s\sigma 3p\pi$     & $D^1\Pi_u 3p\pi$           & $D(3p)^1\Pi_u^\pm$     & $3C^\pm$ & $d^3\Pi_u 3p\pi$           & $d(3p)^3\Pi_u^\pm$    & $3c^\pm$ \\
$1s\sigma 3d\sigma$  & $G^1\Sigma_g^+ 3d\sigma$   & 
                                    \multirow{2}{*}{$GK^1\Sigma_g^+$}      & 
                                                             \multirow{2}{*}{$GK$}    & $g^3\Sigma_g^+ 3d\sigma$   & $g(3d)^3\Sigma_g^+$   & $3d$     \\
                     & $K(^1\Sigma_g^+)$          &                        &          &                            &                       &          \\
$1s\sigma 3d\pi$     & $I^1\Pi_g^ 3d\pi$          & $I(3d)^1\Pi_g^\pm$     & $3E^\pm$ & $i^3\Pi_g 3d\pi$           & $i(3d)^3\Pi_g^\pm$    & $3e^\pm$ \\
$1s\sigma 3d\delta$  &                            & $J(3d)^1\Delta_g^\pm$  & $3F^\pm$ & $j^3\Delta_g 3d\delta$     & $j(3d)^3\Delta_g^\pm$ & $3f^\pm$ \\
                     
$1s\sigma 4p\sigma$  & $B''^1\Sigma_u^+ 4p\sigma$ & $B''(4p)^1\Sigma_u^+$  & $4B$     & $f^3\Sigma_u^+ 4p\sigma$   & $f(4p)^3\Sigma_u^+$   & $4b$     \\
$1s\sigma 4p\pi$     & $D'^1\Pi_u 4p\pi$          & $D'(4p)^1\Pi_u^\pm$    & $4C^\pm$ & $k^3\Pi_u 4p\pi$           & $k(4p)^3\Pi_u^\pm$    & $4c^\pm$ \\
$1s\sigma 4d\sigma$  &                            & $P(4d)^1\Sigma_g^+$    & $4D$     & $p^3\Sigma_g^+ 4d\sigma$   & $p(4d)^3\Sigma_g^+$   & $4d$     \\
$1s\sigma 4d\pi$     &                            & $R(4d)^1\Pi_g^\pm$     & $4E^\pm$ & $r^3\Pi_g 4d\pi$           & $r(4d)^3\Pi_g^\pm$    & $4e^\pm$ \\
$1s\sigma 4d\delta$  &                            &                        &          &                            & $s(4d)^3\Delta_g^\pm$ & $4f^\pm$ \\
                                                                       
$1s\sigma 5p\sigma$  &                            & $B'''(5p)^1\Sigma_u^+$ & $5B$     &                            &                       &          \\
$1s\sigma 5p\pi$     & $D''^1\Pi_u 5p\pi$         & $D''(5p)^1\Pi_u^\pm$   & $5C^\pm$ & $n^3\Pi_u 5p\pi$           & $n(5p)^3\Pi_u^\pm$    & $5c^\pm$ \\
$1s\sigma 5d\sigma$  &                            &                        &          & $q(^3\Sigma_g^+) 5d\sigma$ & $q(5d)^3\Sigma_g^+$   & $5d$     \\
$1s\sigma 5d\pi$     &                            &                        &          & $w(^3\Pi_g) 5d\sigma$      &                       &          \\
                                                                       
$1s\sigma 6p\sigma$  &                            & $(6p)^1\Sigma_u^+$     & $6B$     &                            &                       &          \\
$1s\sigma 6p\pi$     &                            & $(6p)^1\Pi_u^\pm$      & $6C^\pm$ & $u^3\Pi_u 6p\pi$           & $u(6p)^3\Pi_u^\pm$    & $6c^\pm$ \\
$1s\sigma 6d\sigma$  &                            &                        &          &                            & $(6d)^3\Sigma_g^+$    & $6d$     \\
                                         
$1s\sigma 7p\sigma$  &                            & $(7p)^1\Sigma_u^+$     & $7B$     &                            &                       &          \\
$1s\sigma 7p\pi$     &                            & $(7p)^1\Pi_u^\pm$      & $7C^\pm$ &                            & $(7p)^3\Pi_u^\pm$     & $7c^\pm$ \\
$1s\sigma 7d\sigma$  &                            &                        &          &                            & $(7d)^3\Sigma_g^+$    & $7d$     \\
                                         
$1s\sigma 8p\sigma$  &                            & $(8p)^1\Sigma_u^+$     & $8B$     &                            &                       &          \\
$1s\sigma 8p\pi$     &                            & $(8p)^1\Pi_u^\pm$      & $8C^\pm$ &                            & $(8p)^3\Pi_u^\pm$     & $8c^\pm$ \\
$1s\sigma 8d\sigma$  &                            &                        &          &                            & $(8d)^3\Sigma_g^+$    & $8d$     \\
                                         
$1s\sigma 9p\sigma$  &                            & $(9p)^1\Sigma_u^+$     & $9B$     &                            &                       &          \\
$1s\sigma 9p\pi$     &                            & $(9p)^1\Pi_u^\pm$      & $9C^\pm$ &                            & $(9p)^3\Pi_u^\pm$     & $9c^\pm$ \\
$1s\sigma 9d\sigma$  &                            &                        &          &                            & $(9d)^3\Sigma_g^+$    & $9d$     \\
                     
\hline
\end{longtable}
}

Two different notations are used for the electronic states of the $D_2$ \cite{Huber, FSC1985}. 
Traditional notation \cite{Huber} does not need any additional explanations. It is based on 
the assumption of the adiabatic approximation and Hund's case 'b' for angular momenta 
coupling. The notation earlier introduced by G.H.Dieke \cite{Dieke1972} (and later made 
more exact in \cite{FSC1985}) is based on the same assumptions but it is much more 
compact what is very important for long tables of spectral lines. For example, the 
$(s\sigma)\Sigma_g^+$ states are denoted by uppercase letter "A" for singlets and 
by lowercase letter "a" for triplets, the $(p\sigma)\Sigma_u^+$ states are marked 
by "B" or "b", the $(p\pi)\Pi_u$ states by "C" or "c", the $(d\sigma)\Sigma_g^+$ 
states by "D" or "d" etc. with the principal quantum number $n$ of the excited 
electron for united atom limit case is included as a prefix.

The relation between notations from \cite{Huber} and \cite{FSC1985}
is presented in Table~\ref{notations}. One may see that Dieke's notations are
much more compact than traditional. 
Only part of all investigated electronic states of $D_2$ is listed in
Table~\ref{notations}. Singlet states $(np)^1\Sigma_u^+$ up to $n=46$ were
studied in absorption in the VUV region of the $D_2$ spectrum \cite{TT1975},
the study of these states is outside the scope of this paper and they are not
listed in Table~\ref{notations}.

For the rovironic transitions Dieke's notation consists of: the multiplicity (all lines
are assigned to either singlet-to-singlet of a triplet-to-triplet transition, indicated 
by S or T respectively), the reflection (Kr\"onig) symmetry of the upper level (is given
as $+$ or $-$), electronic transition in Dieke's notations with the upper level coming
first, vibrational quantum numbers of the transition in the parentheses with the upper state listed 
first and finally the rotational branch ($P$, $Q$, or $R$) followed by the rotational 
quantum number $N''$ of the lower rovibronic state.
For example the notation "T+  4b-2a  (2-3) P1" indicates the transition between
$f(4p)^3\Sigma_u^+$, $v'=2$, $N'=0$ and $a(2s)^3\Sigma_g^+$, $v''=3$, $N''=1$ rovibronic 
levels in traditional notation.

\section*{Experimetnal}

We used experimental setup described elsewhere \cite{ALMU2008, LMU2011}. 
Emission of plasma inside molybdenum capillary located between anode and 
cathode of a gas discharge tube was used as a light source. The flux of 
radiation through a hole in an anode was focused by achromatic lens on the 
entrance slit of the spectrometer. Detailed description of the self-made high 
resolution automatic spectrometer and corresponding software was reported in 
\cite{LMU2011}. The $2.65$ m Ebert-Fastie spectrograph with $1800$ grooves per 
mm diffraction grating was equipped with additional camera lens (that gives 
effective focus length $F=6786 \pm 8$ mm) and computer-controlled CMOS matrix 
detector ($22.2 \times 14.8$ mm$^2$, $1728 \times 1152$ pixels). The apparatus 
has linear dispersion of $0.076 \div 0.065$ nm/mm in the wavelength range 
$400-700$ nm, dynamic range of measurable intensities greater than $10^4$ and 
maximum resolving power up to $2 \times 10^5$. However, actual resolving power 
in our conditions was mainly limited by Doppler broadening of the $D_2$ 
spectral lines due to small reduced mass of nuclei.

For recording the $D_2$ spectra with moderate resolution and large population
of high rotational levels we used hot-cathode capillary-arc discharge
lamp LD-2D described in \cite{GLT1982} (pure $D_2$ under pressure 
$\approx 6$ Torr, capillary inner diameter \O{2} mm, current 
density $\approx 10$ A/cm$^2$). Gas temperature $T = 1890 \pm 170 $ K
was obtained from the intensity distribution in the rotational structure of
the $(2-2)$ Q-branch of Fulcher-$\alpha$ band system (see e.g. 
\cite{L1980, AKKK1996}). It corresponds to Doppler linewidths (FWHM) 
$\Delta \nu_D$ = $0.22 \div 0.37$ cm$^{-1}$ for $1/\nu = 420 \div 700$ nm.
Therefore we were able to open the entrance slit of the spectrometer up to 
60 $\mu$m for gaining more signal (and corresponding decrease in data 
accumulation time) without significant loss in resolution.

Our way of determination of the rovibronic transition wavenumbers developed in 
\cite{ALMU2008, LU2009, LU2008, LMU2011, LUZ2012} is based on linear response 
of the CMOS matrix detector on the spectral irradiance and digital intensity 
recording. Both things provide an extremely important
advantage of our technique over traditional photographic recording 
with microphotometric comparator reading. It
not only makes it easier to measure the relative spectral line
intensities but also makes it possible to investigate the shape
of the individual line profiles and, in the case of overlap of
the contours of adjacent lines (so-called blending), to carry
out numerically the deconvolution operation (inverse to the convolution
operation) and thus to measure the intensity and wavelength
of even blended lines. 

It is known that, in the case of long-focus spectrometers, the dependence of 
the wavelength on the coordinate $x$ along direction of dispersion, is close 
to linear in the vicinity of the center of the focal plane. It can be 
represented as a power series expansion over of the small parameter $x/F$, 
which in our case does not exceed $2 \times 10^{-3}$.\footnote{The 
$x$-coordinate actually represents small displacement from the center 
of the matrix detector. $F$ is the focal length of the spectrometer mirror.}
On the other hand, the wavelength dependence of the refractive index of air 
$n(\lambda)$ is also close to linear inside a small enough part of the 
spectrum. Thus, when recording narrow spectral intervals, the product 
$\lambda_{vac}(x) = \lambda(x) n(\lambda(x))$ has the form of a power series 
of low degree. This circumstance makes it possible to calibrate the 
spectrometer directly in vacuum wavelengths $\lambda_{vac} = 1 / \nu$, thereby 
avoiding the technically troublesome problem of accurate measuring the 
refractive index of air under the various conditions under which measurements 
are made. 

Another peculiarity of our calibration technique is using of experimental 
vacuum wavelength values from \cite{FSC1985} as standard reference data. 
For bright unblended spectral lines the wavelength values show small 
random spread around smooth curve approximating the dependence of the 
wavelengths of the lines against their positions in the focal 
plane of the spectrometer. Moreover those random 
errors are in good accordance with normal Gaussian distribution function. Thus 
it is possible to obtain precision for new wavenumber values better than that 
of the reference data due to smoothing. 

To be sure that the data from \cite{FSC1985} are free from systematic errors we have had to perform 
special experiments with capillary-arc lamp analogous to that described in \cite{LSh1979} 
(capillary diameter $d = 1.5$ mm and current density $j = 30$ A/cm$^2$) but filled with the $H_2+D_2+Ne$ 
mixture (1:1:2) under total pressure $P \approx 8$ Torr \cite{LU2011}.

\begin{figure}[!ht]
\begin{center}
\epsfig{file=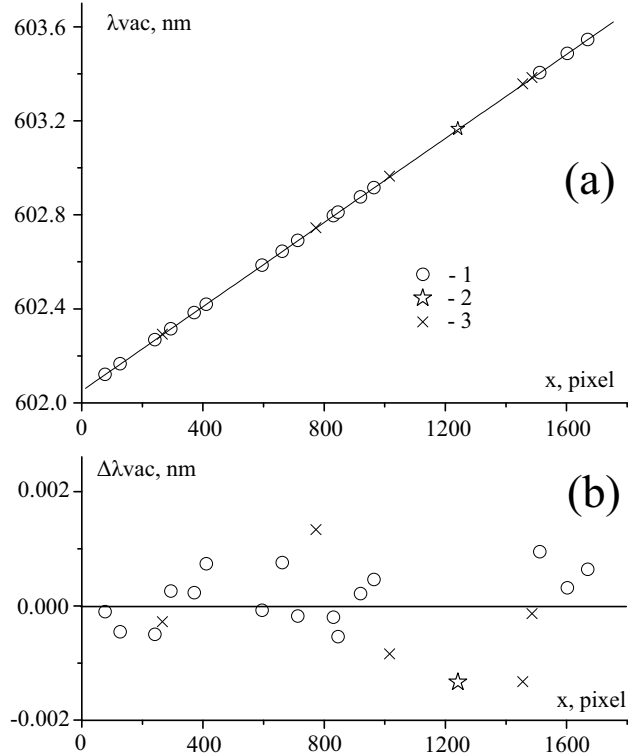, width=0.5\columnwidth,clip}
\end{center}
\caption{Dependences of vacuum wavelengths $\lambda_{vac}$ of the brightest 
$D_2$, $H_2$ and $Ne$ spectral lines on the coordinate (in pixels) in the focal plane 
of the spectrometer (a) and their deviations $\Delta \lambda_{vac}$ from 
the callibration curve (b); 
$1$ are the values for the $D_2$ molecule from \cite{FSC1985},
$2$ --- for the $Ne$ atom from \cite{SS2004},
$3$ --- for the $H_2$ molecule from \cite{Dieke1972};
solid line represent the approximation of experimental data. \cite{LU2011}}\label{d2h2ne}
\end{figure}

For vacuum wavelength calibration we used bright free of blending lines of the $D_2$ 
and $H_2$ molecules as well as $Ne$ spectral lines with reference data from \cite{FSC1985, Dieke1972, SS2004} 
respectively. 
The dependence of vacuum line wavelength
on its position on CMOS matrix in pixels for strong unblended lines is shown on fig.\ref{d2h2ne}(a).
One may see that the dependence of the wavelengths for most of the lines on the coordinate is
monotonic and close to linear. 
The calibration curve of the spectrometer was obtained by the polynomial least-squares fitting of the data.
Our measurements showed that, using a linear hypothesis is inadequate and a third--degree polynomial
is excessive, while an approximation by a second-degree polynomial provides calibration accuracy 
better than $2 \times 10^{-3}$ nm.
Such a wavelength calibration allows us to get new experimental values for the rovibronic line wavenumbers.
The differences $\Delta \lambda_{vac}$ between the new values and the used reference data are shown 
in fig.\ref{d2h2ne}(b).
One may see that the differences have certain spread around calibration curve, that does not exceed $0.002$ nm.
Thus our measurements show that experimental wavenumber values from \cite{FSC1985, Dieke1972, SS2004} 
are in good agreement with each other. 
Therefore in our studies of the $D_2$ spectrum the vacuum wavelengths values from \cite{FSC1985} were used as
the reference data set.
Such "internal reference light source" gave us an opportunity to eliminate 
experimental errors caused by the shift between a spectrum under 
the study and the reference spectrum from another reference light source, due 
to a different illumination of the grating by the different lamps 
(see e.g.\cite{RLTBjcp2006}). 

Each experimental wavenumber value measured in the framework of procedure 
described above is obtained with the uncertainty (one standard deviation) 
determined by the quality of approximation of a recorded intensity distribution 
by a sum of spectral line profiles and the quality and quantity of 
standard reference data within every selected for processing small fragment 
of the spectrum. 

\section*{Results and discussion}

The visible part ($\approx 419 \div 696$ nm) of the emission spectrum of 
the $D_2$ plasma was recorded and analyzed by means of technique described above.
In atlas and tables of the present work we report part of this spectrum,
namely the wavelength region ($\approx 419 \div 550$ nm).
Second part of this spectrum ($550 \div 696$ nm) will be reported in the subsequent paper.
It contains two lines of the atomic deuterium ($D_\beta$, and $D_\gamma$),
corresponding lines of atomic hydrogen (impurity) and 6541 lines of molecular deuterium.
The results are reported in the form of the
atlas divided into 36 sections each covering about 1.5 nm, containing pictures of 
images in the focal plane of the spectrometer, intensity distributions in 
linear and logarithmic scales and the table containing wavenumber and relative 
intensity values for recognized spectral lines together with existing line assignments.
Positions of spectral lines obtained by the deconvolution are presented as "stick diagrams" 
indicating their wavenumbers and amplitudes.
The numbering of the lines (for every fifth line) is shown under the 
intensity distributions in linear scale.

All measured wavenumber values for assigned triplet spectral lines were 
used for obtaining the set of optimal rovibronic energy levels using the method
of statistical analysis \cite{LRJetf2005} with experimental data \cite{FSC1985, 
DabrHerz, Davies, DiekeBlue1935, Dieke1935_2, GloersenDieke, DiekePorto, FMZ1976}. 
Detailed description of the analysis will be provided elsewhere. 
It was carried out similar to our previous work \cite{LU2008}, but
the observation of pseudo doublets \cite{LUZ2012} forced us to carry out 
the optimization in two stages. At the first stage spectral line wavenumber 
values for band systems having one common low electronic state $a^3\Sigma_g^+$ 
($n^3\Lambda_g - a^3\Sigma_g^+$, with $\Lambda = 0, 1$ and $n = 3 - 9$) were
analyzed. Obtained values of rovibronic energy levels were fixed and then all
other wavenumber values were added to the optimization procedure. Such a 
two-stage procedure gave us opportunity to obtain 595 energy level values of 
$a^3\Sigma_g^+$, $n^3\Lambda_g$ with $\Lambda = 0, 1$ and $n = 3 - 9$
electronic states having small fine structure splitting value with high 
precision. The values for 450 energy level values of 
$c^3\Pi_u$, $n^3\Lambda_u$ with $\Lambda = 0, 1$ and $n = 3 - 9$
electronic states are less accurate due to observed spectral lines fine 
structure.
Our statistical analysis shows good agreement in the framework of the 
Rydberg-Ritz principle between used wavenumbers of spectral lines spread over 
the very wide range 0.896--28166.84 cm$^{-1}$ from radio frequencies up to 
the ultraviolet obtained for various band systems, by various methods and 
authors, and in various works.

Table contains: first column --- spectral line number $K$, 
second and third column --- measured wavenumber $\nu$ and 
intensity $I$ values respectively with standard deviation 
in units of last significant digit, fourth column --- wavenumber value 
of the line from \cite{FSC1985} in the cases when it was used as a 
reference data and the fifth column --- assignment in the Dieke's notations.
Confirmed by statistical analysis assignments for triplet lines are 
shown in bold and the new assignments are shown in italic.

\begin{acknowledgments}
Present work was supported, in part, by 
the Russian Foundation for Basic Research, Grant No. 10-03-00571-a.
\end{acknowledgments}

\newpage

\begin{landscape}
{\def\baselinestretch{1.0}
\footnotesize		
\setlength{\tabcolsep}{1pt}

\begin{longtable}[]{r|lr|l|r|r|lr|l|r}
\caption[]{$D_2$ rovibronic spectral lines vacuum wavenumbers values, obtained in present work $\nu$.
The uncertainties of the $\nu$ value determination (one SD) are shown in brackets
in units of last significant digit.
$\nu_R$ --- wavenumber from \cite{FSC1985} used as reference data.}\label{tabnewlines01}
\endfirsthead

\multicolumn{10}{l}{Table~\ref{tabnewlines01} (Continued).}\\ 
\hline
K & $\nu$, cm$^{-1}$ & $I$, counts & $\nu_R$, cm$^{-1}$ & Assignment &
K & $\nu$, cm$^{-1}$ & $I$, counts & $\nu_R$, cm$^{-1}$ & Assignment \\
\hline
\endhead

\hline
K & $\nu$, cm$^{-1}$ & $I$, counts & $\nu_R$, cm$^{-1}$ & Assignment &
K & $\nu$, cm$^{-1}$ & $I$, counts & $\nu_R$, cm$^{-1}$ & Assignment \\
\hline

\input{NewLineTab01.tex}

\hline
\end{longtable}
}

\newpage

\end{landscape}

\input{LU2012xxx_atl.tex}

\end{document}

%% file: LU2012xxx_atl.tex
\begin{figure}[!ht]
\includegraphics[angle=90, totalheight=0.9\textheight]{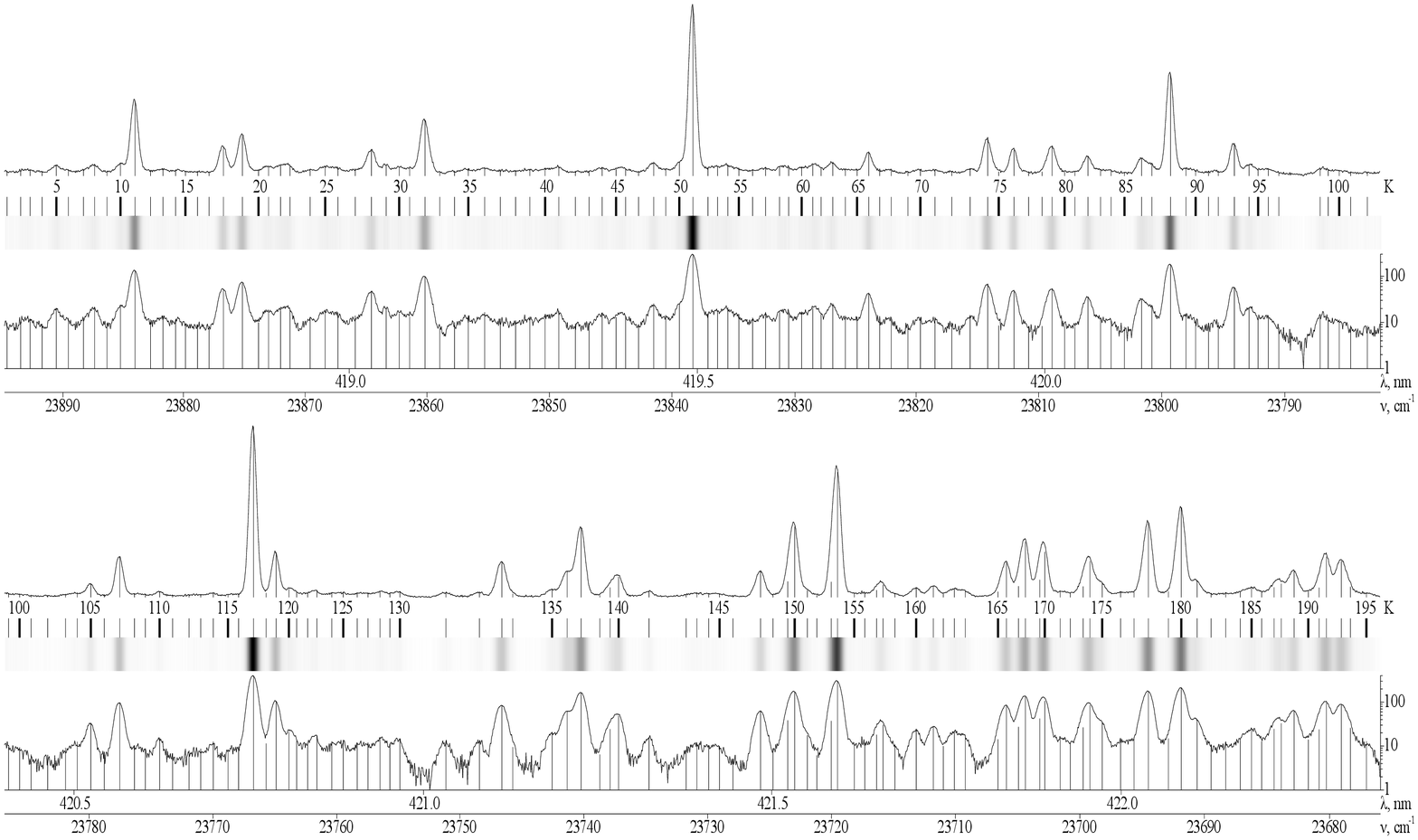}
\end{figure}

\newpage
\begin{figure}[!ht]
\includegraphics[angle=90, totalheight=0.9\textheight]{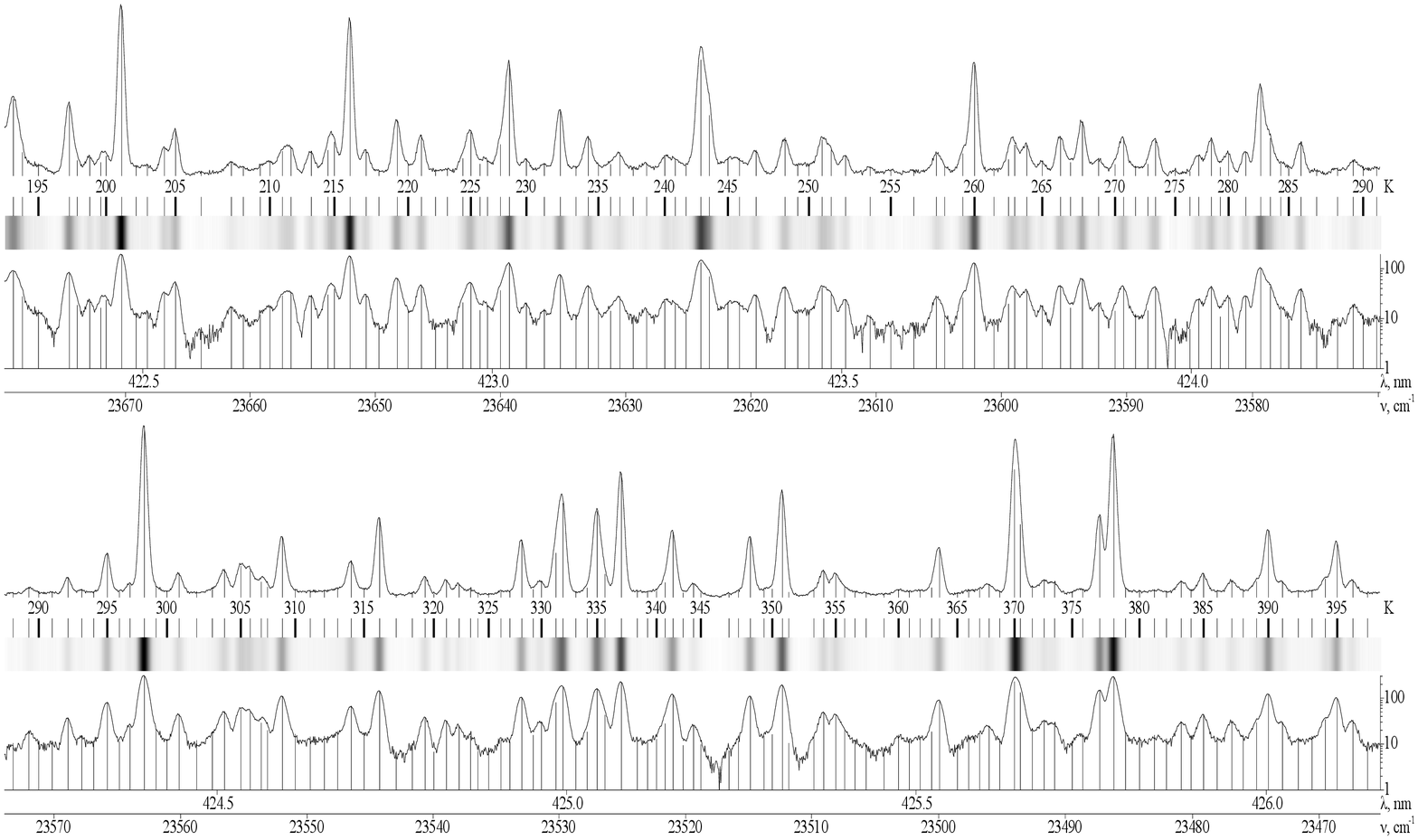}
\end{figure}

\newpage
\begin{figure}[!ht]
\includegraphics[angle=90, totalheight=0.9\textheight]{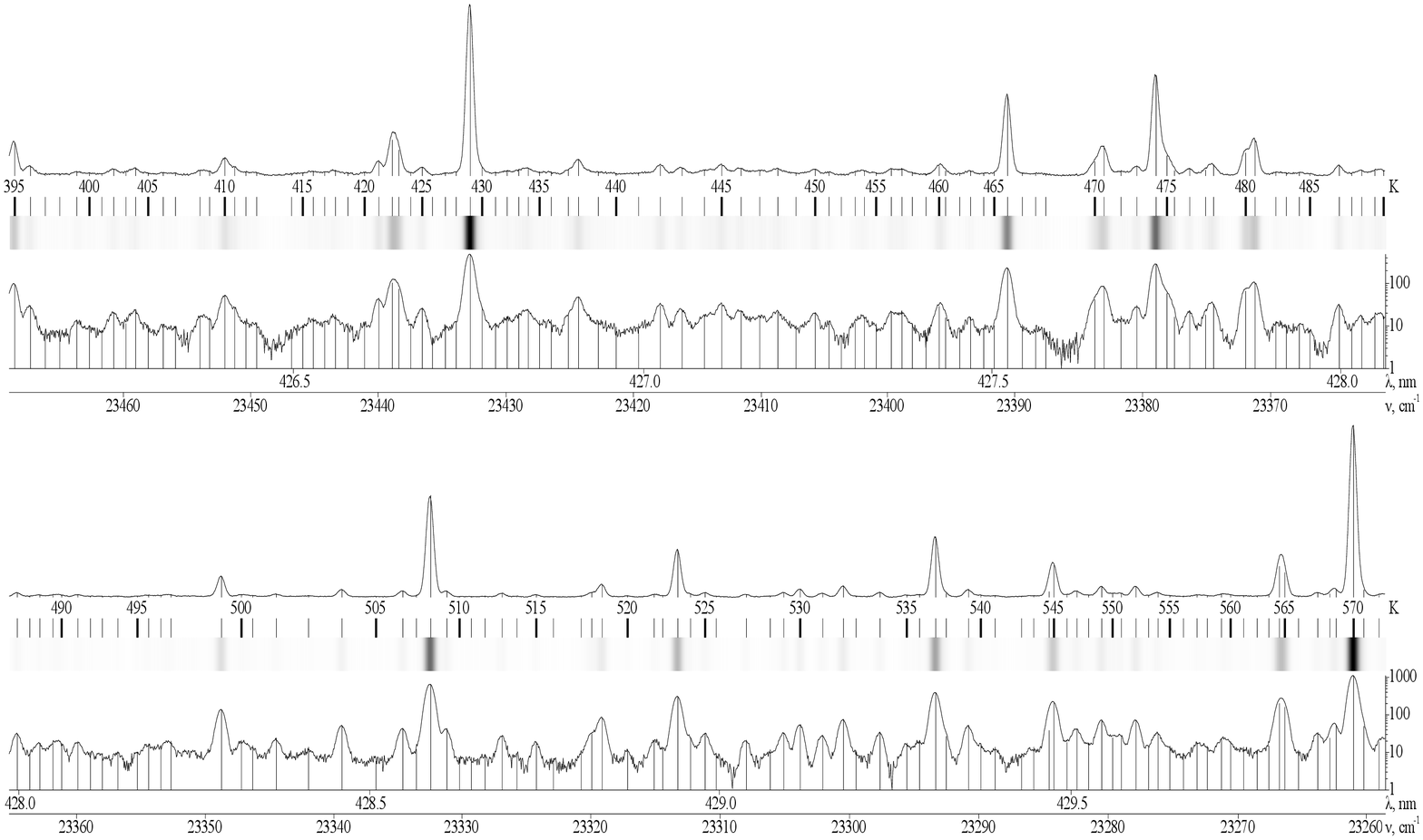}
\end{figure}

\newpage
\clearpage
\begin{figure}[!ht]
\includegraphics[angle=90, totalheight=0.9\textheight]{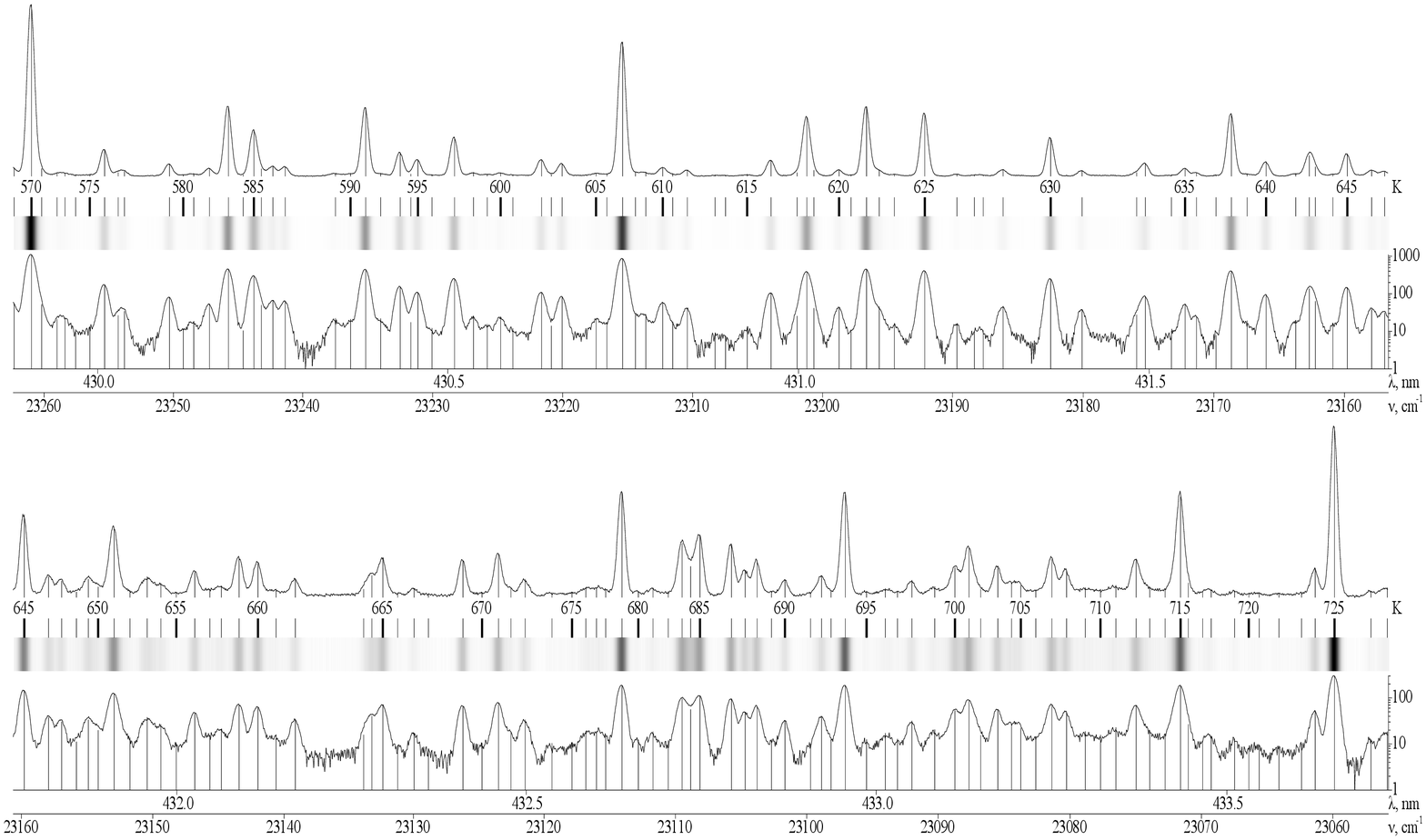}
\end{figure}

\newpage
\begin{figure}[!ht]
\includegraphics[angle=90, totalheight=0.9\textheight]{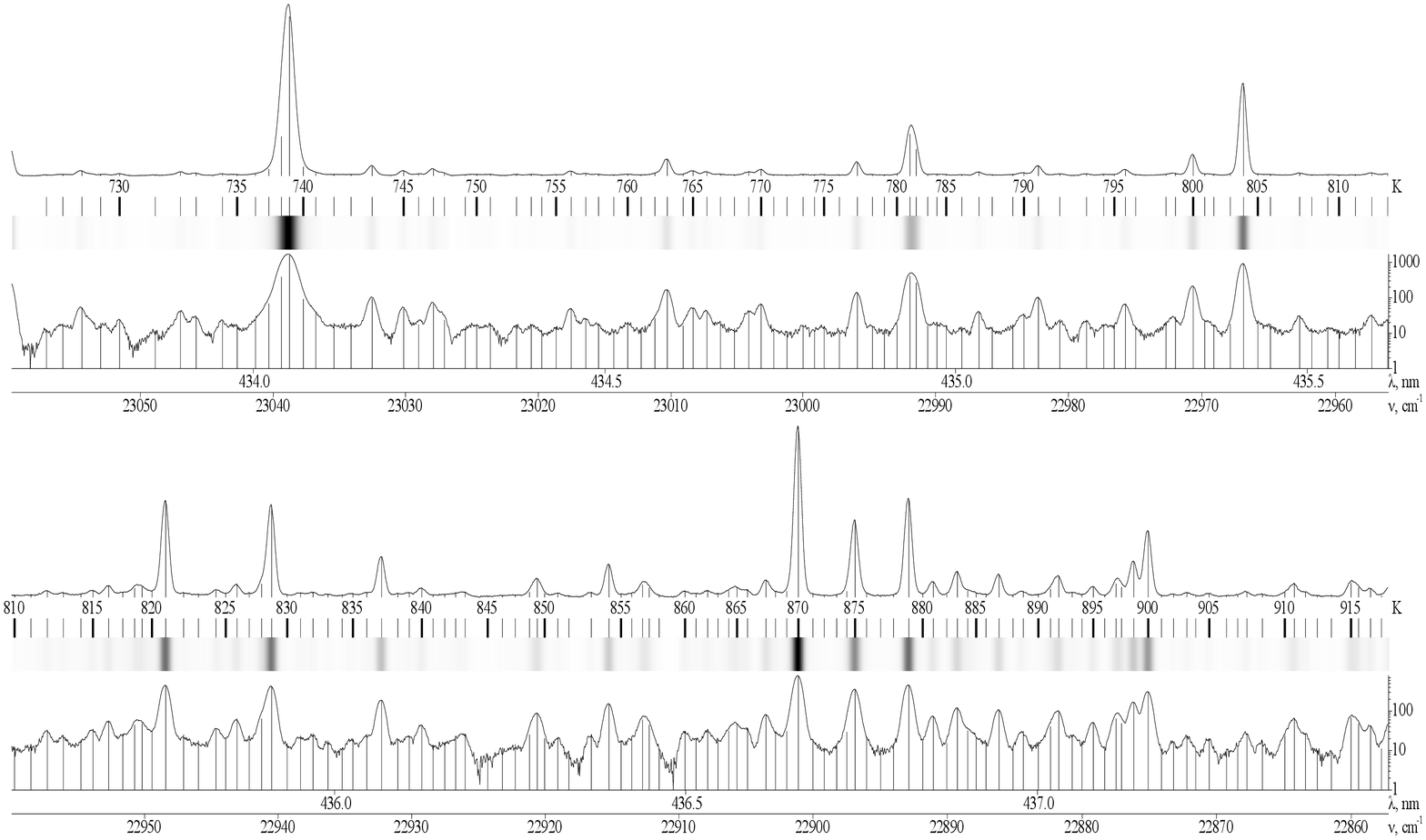}
\end{figure}

\newpage
\begin{figure}[!ht]
\includegraphics[angle=90, totalheight=0.9\textheight]{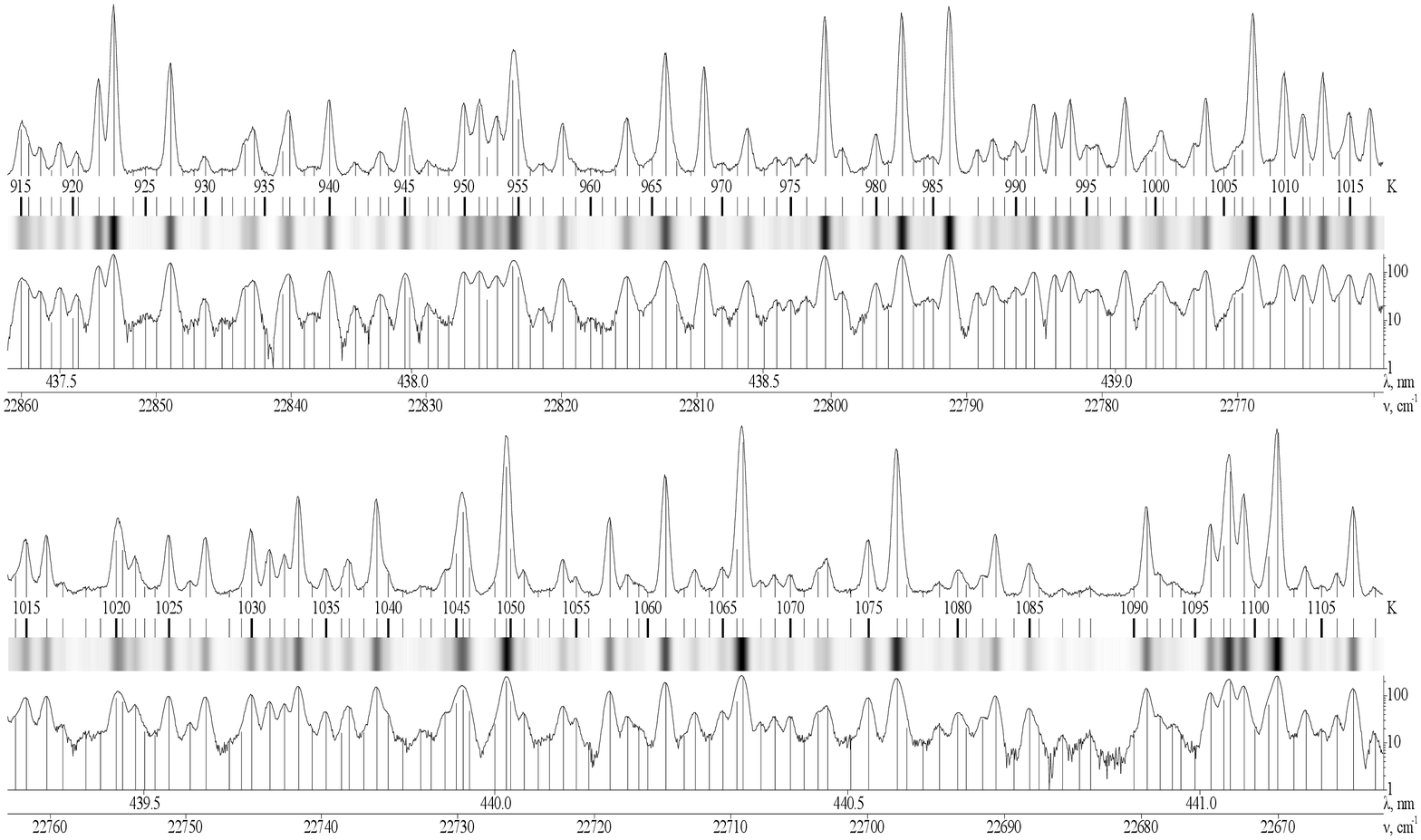}
\end{figure}

\newpage
\begin{figure}[!ht]
\includegraphics[angle=90, totalheight=0.9\textheight]{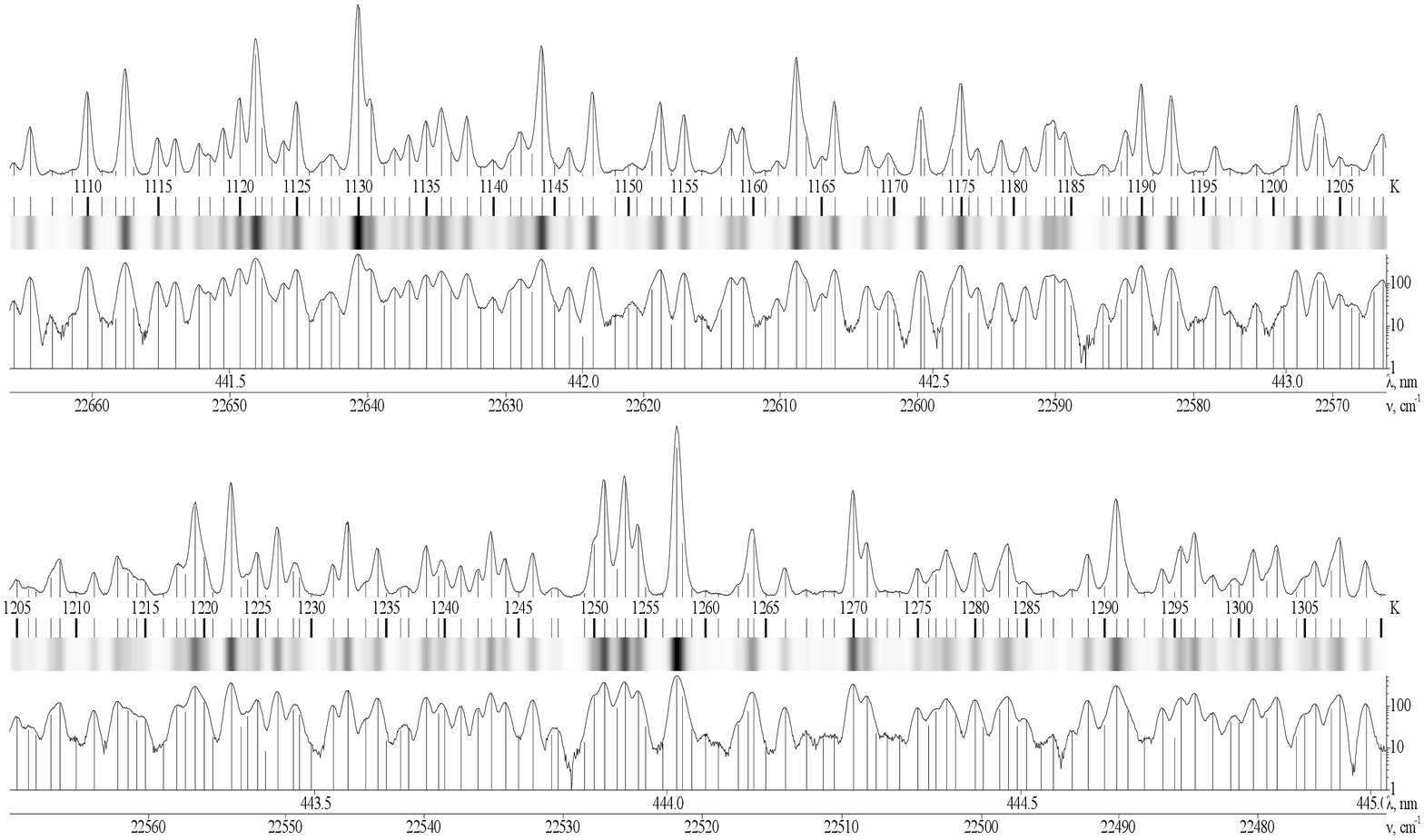}
\end{figure}

\newpage
\begin{figure}[!ht]
\includegraphics[angle=90, totalheight=0.9\textheight]{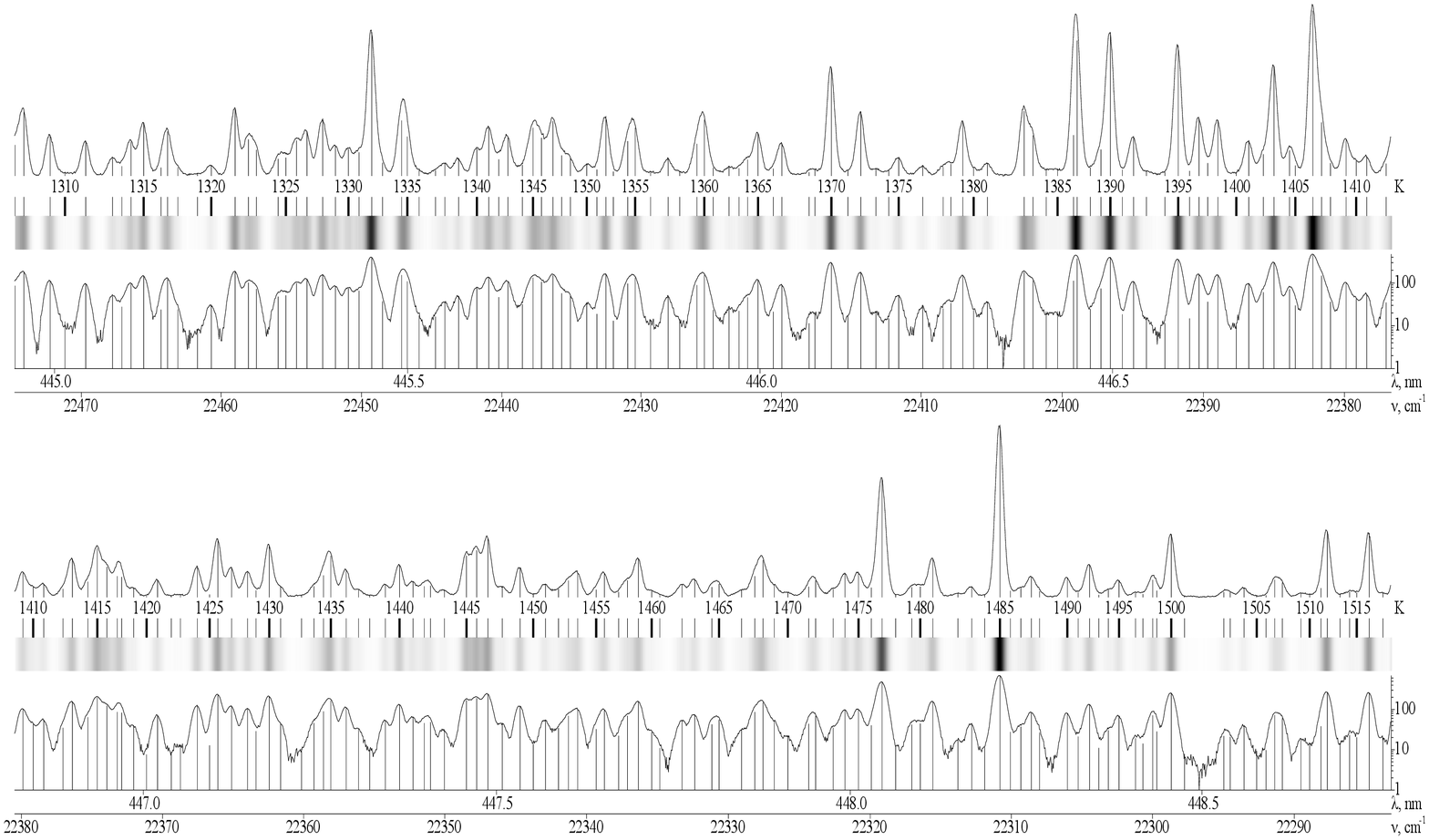}
\end{figure}

\newpage
\begin{figure}[!ht]
\includegraphics[angle=90, totalheight=0.9\textheight]{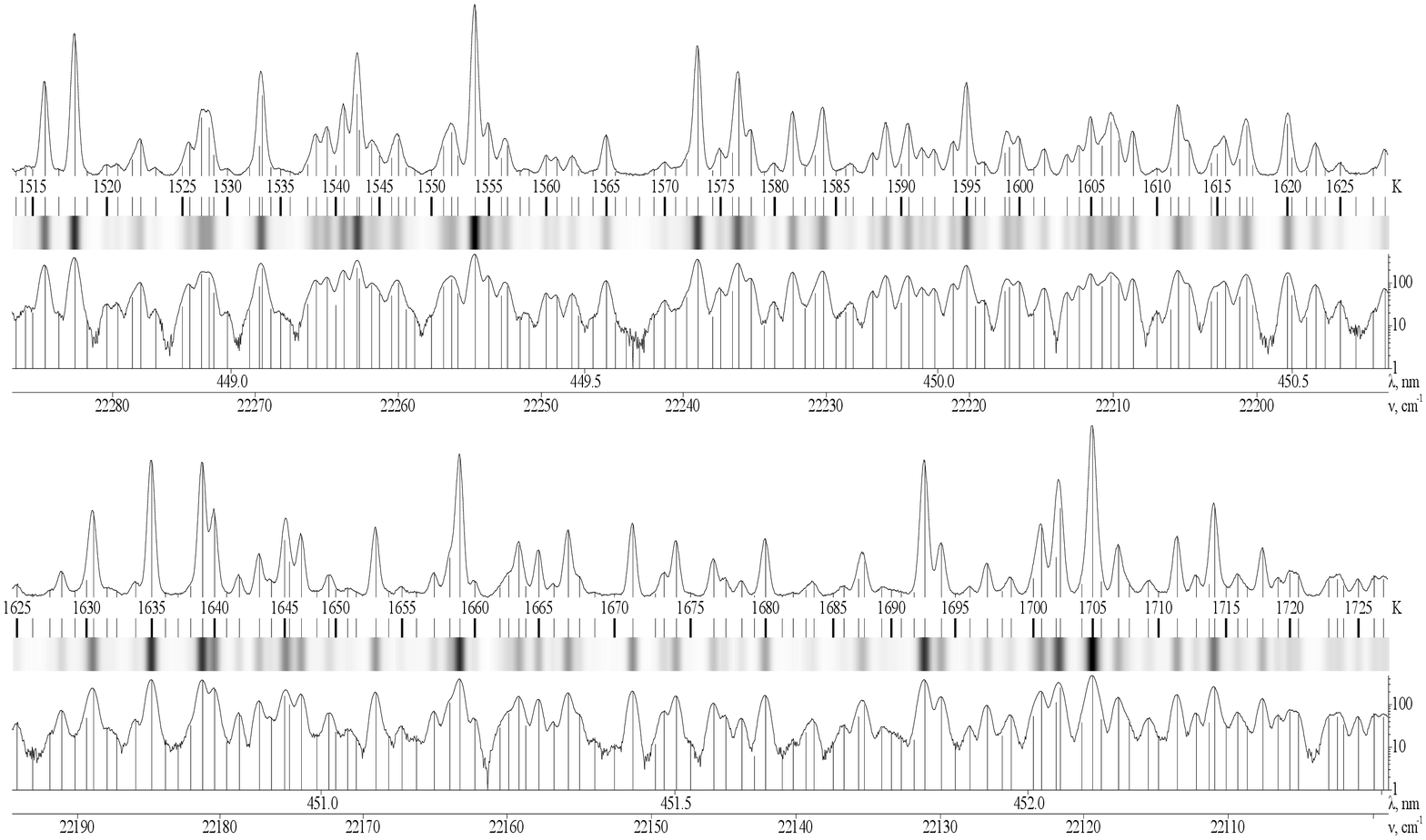}
\end{figure}

\newpage
\begin{figure}[!ht]
\includegraphics[angle=90, totalheight=0.9\textheight]{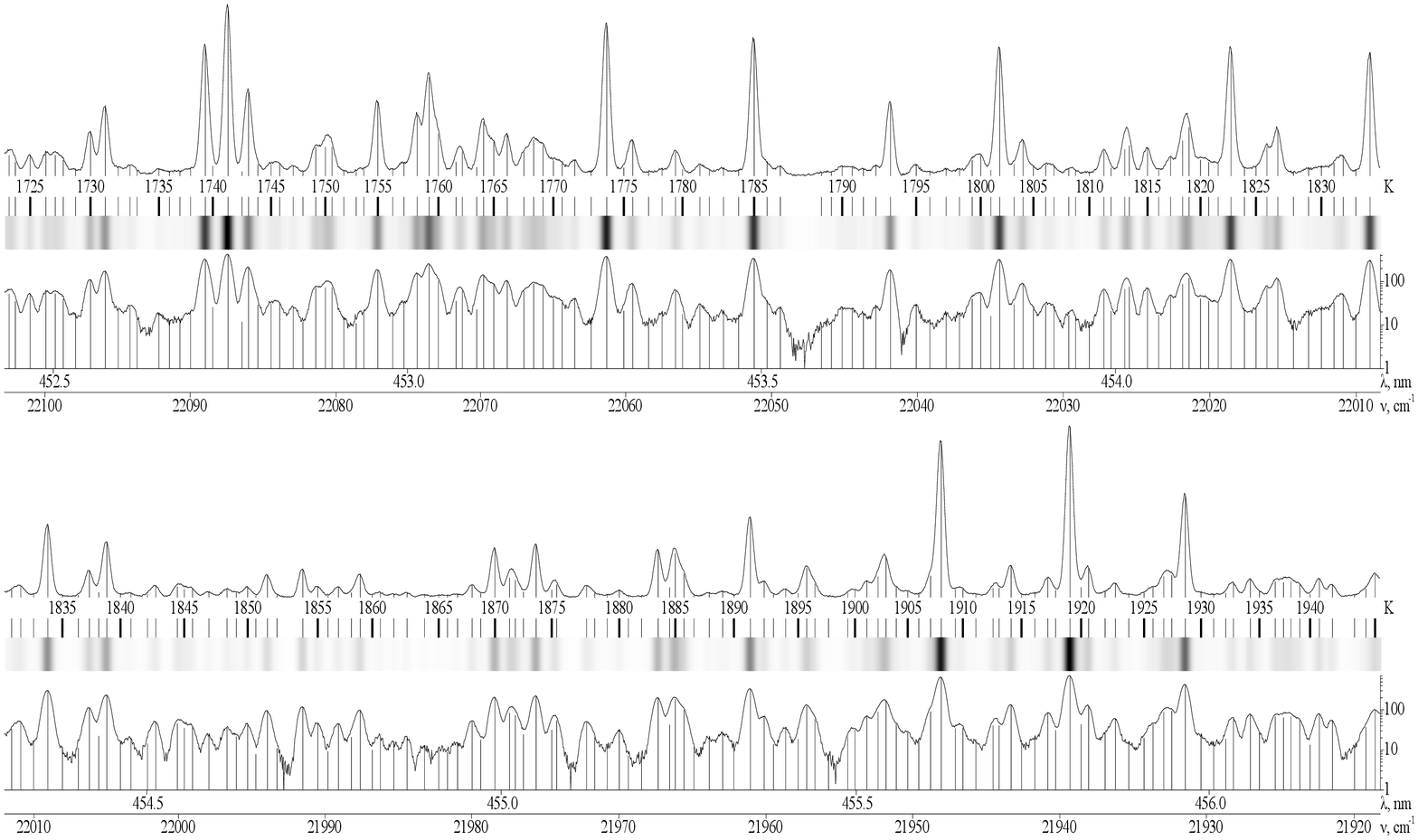}
\end{figure}

\newpage
\begin{figure}[!ht]
\includegraphics[angle=90, totalheight=0.9\textheight]{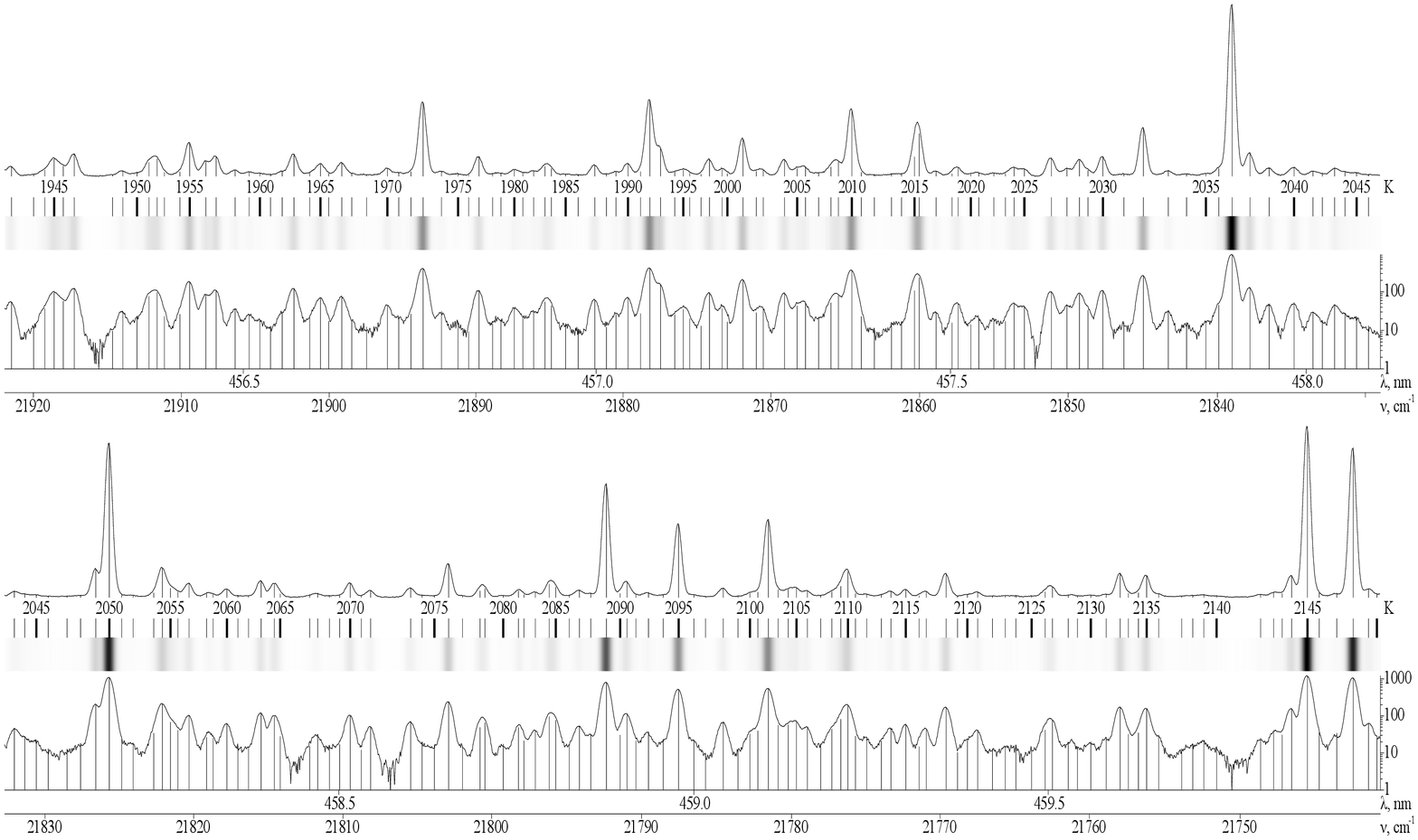}
\end{figure}

\newpage
\begin{figure}[!ht]
\includegraphics[angle=90, totalheight=0.9\textheight]{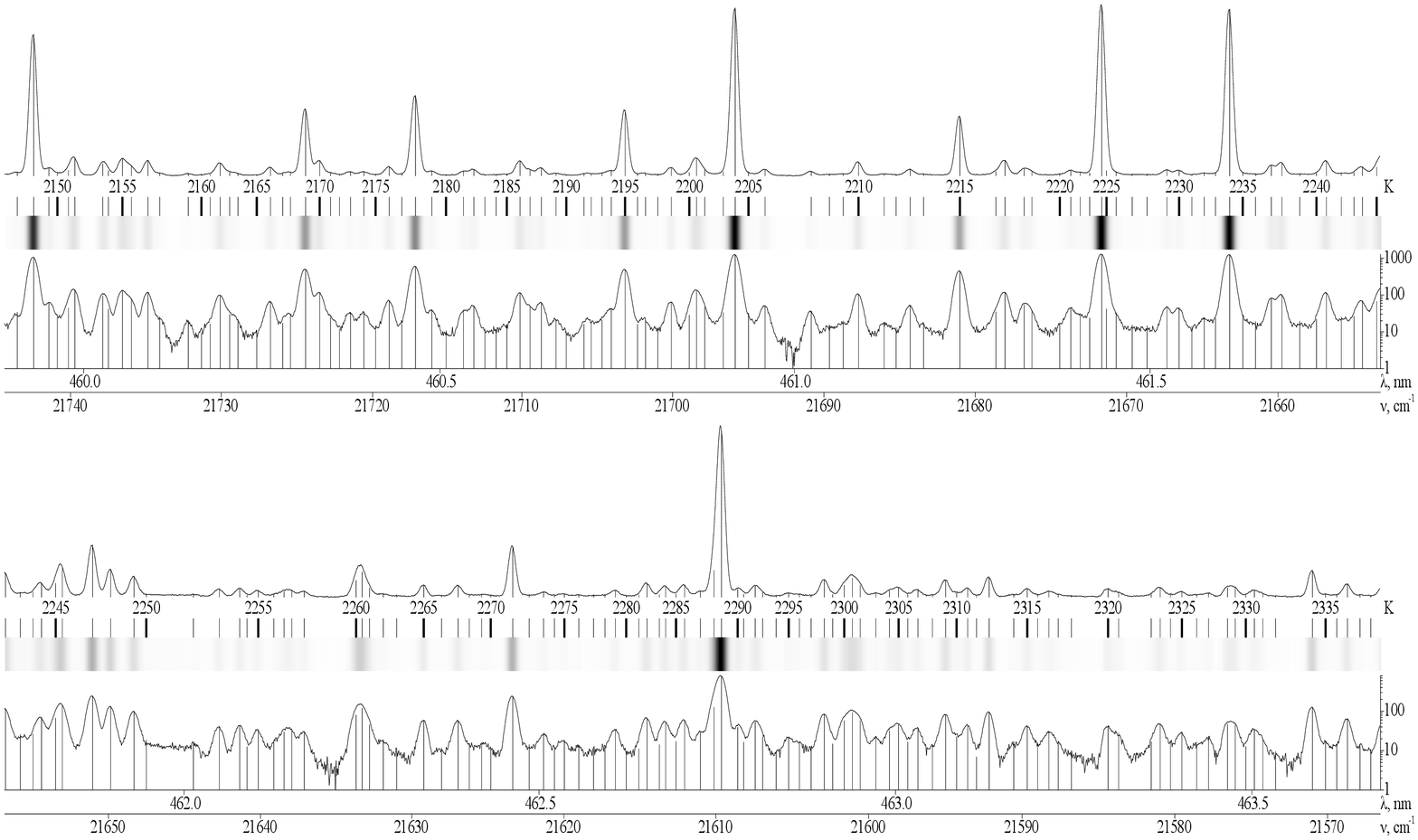}
\end{figure}

\newpage
\begin{figure}[!ht]
\includegraphics[angle=90, totalheight=0.9\textheight]{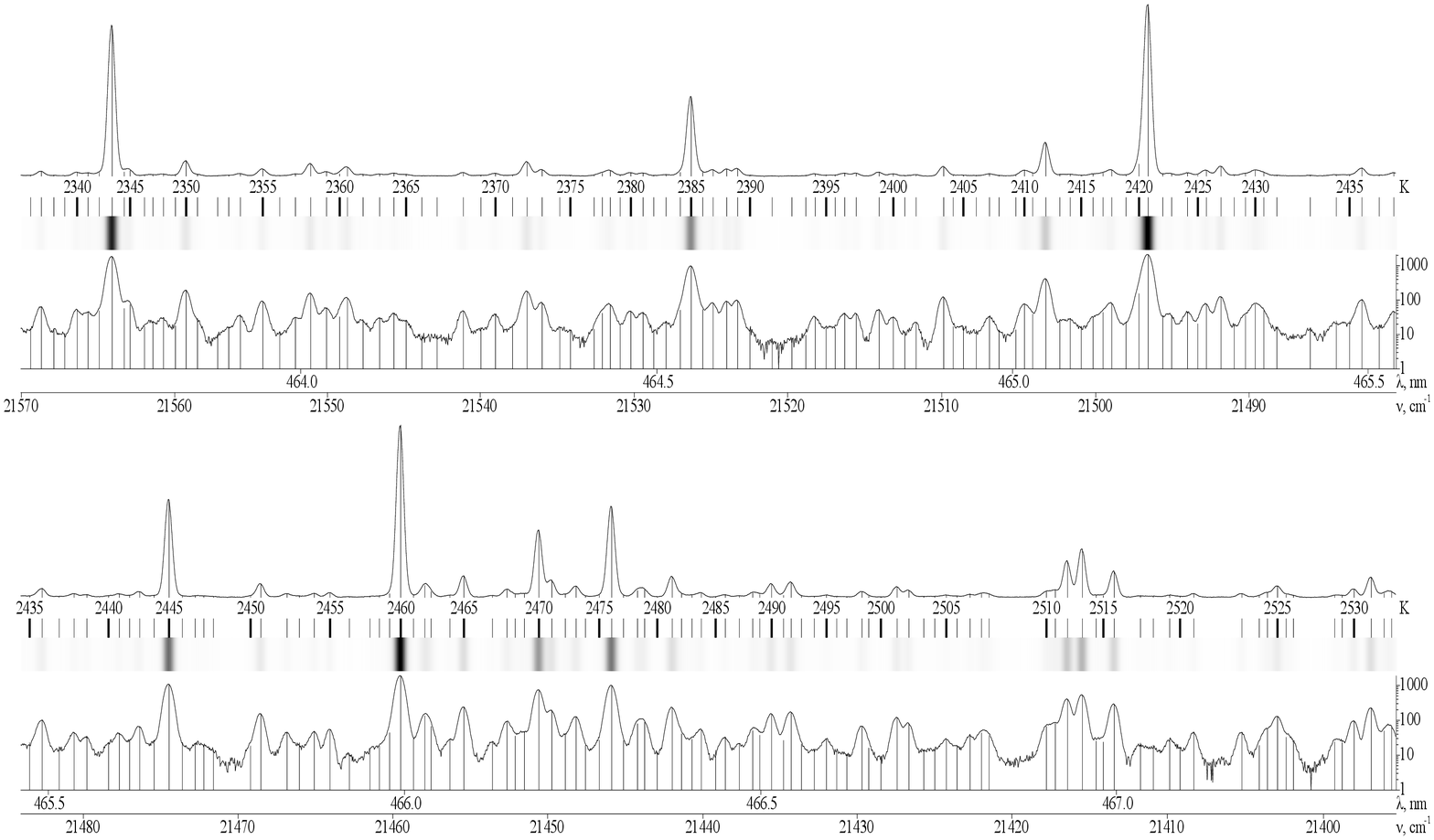}
\end{figure}

\newpage
\begin{figure}[!ht]
\includegraphics[angle=90, totalheight=0.9\textheight]{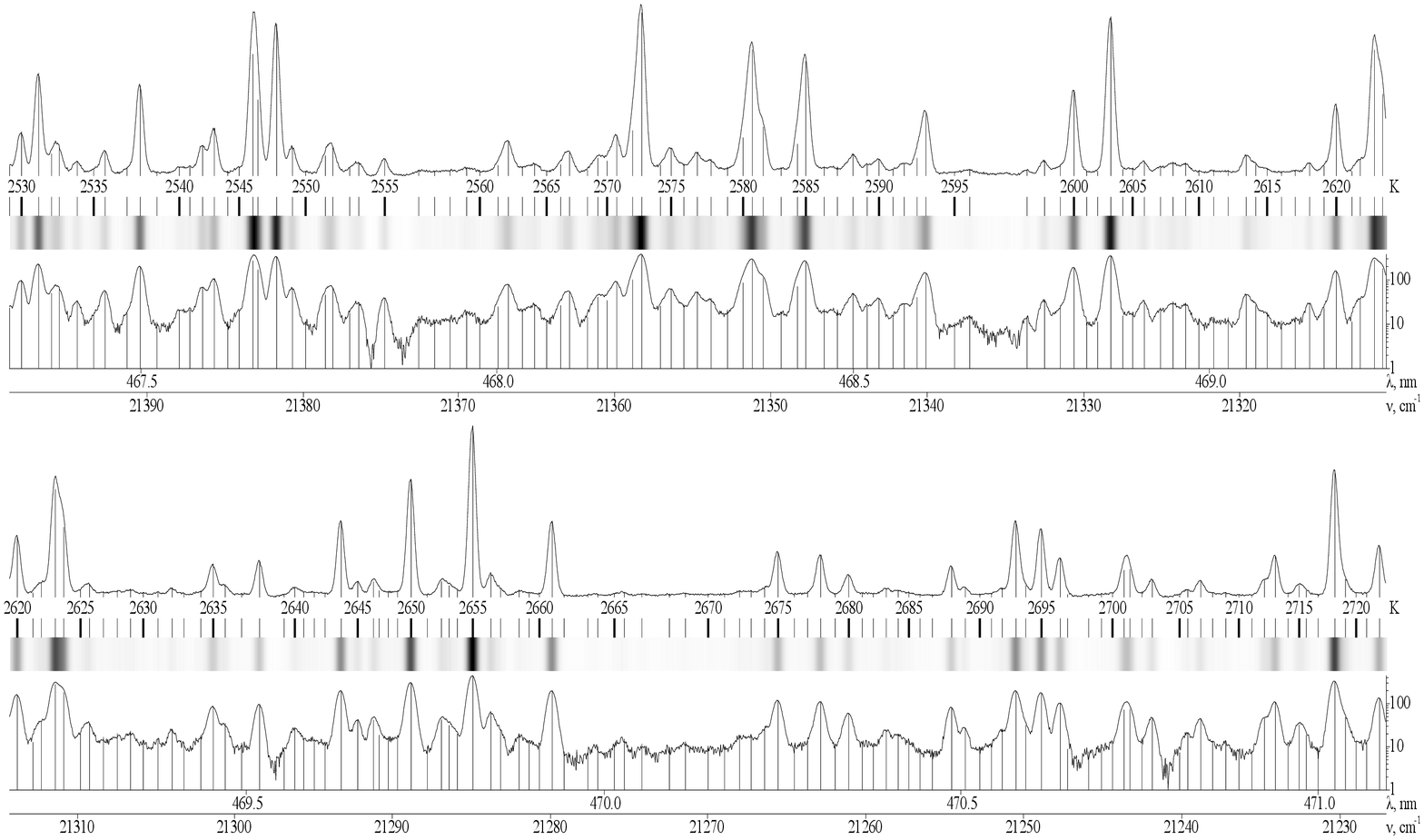}
\end{figure}

\newpage
\begin{figure}[!ht]
\includegraphics[angle=90, totalheight=0.9\textheight]{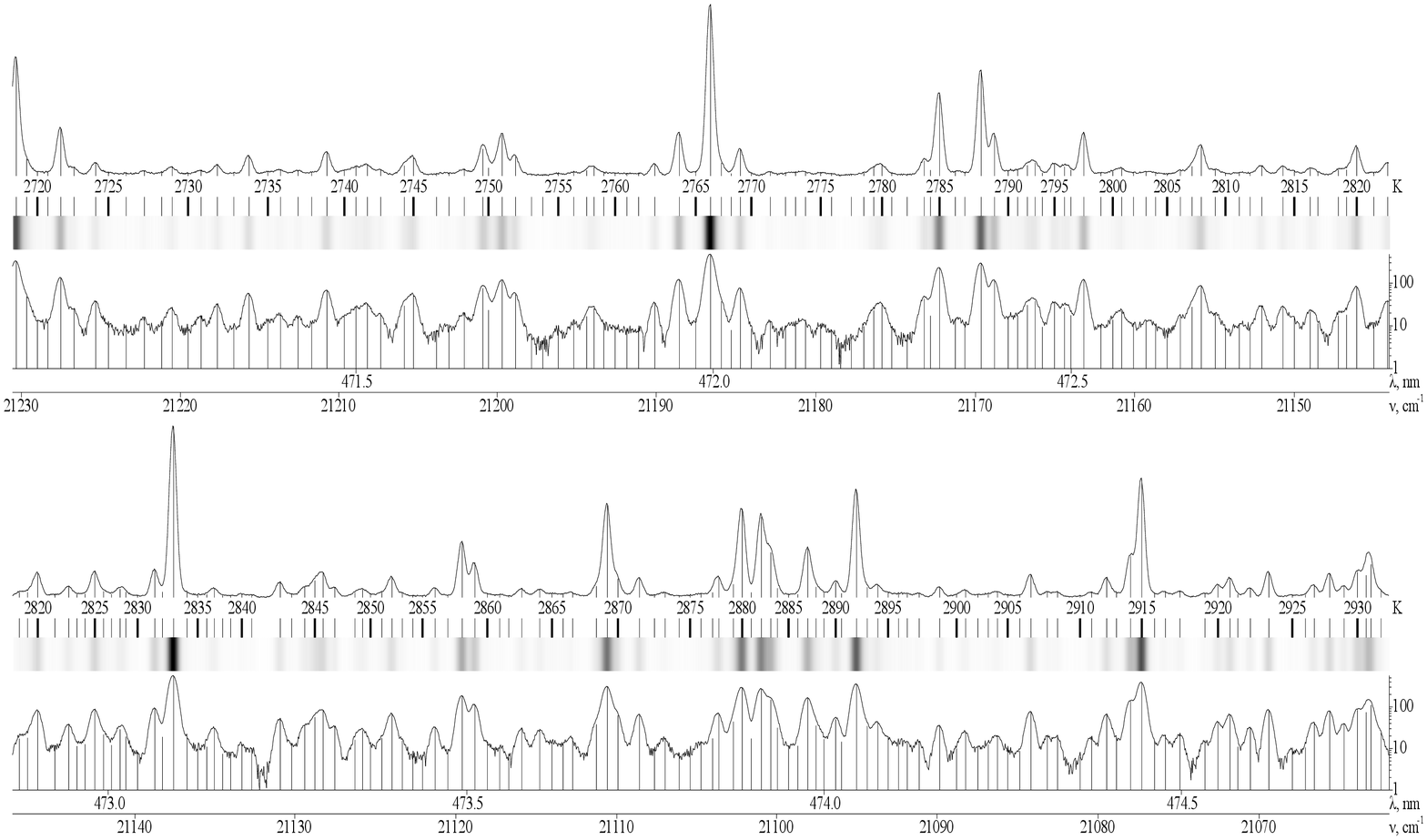}
\end{figure}

\newpage
\begin{figure}[!ht]
\includegraphics[angle=90, totalheight=0.9\textheight]{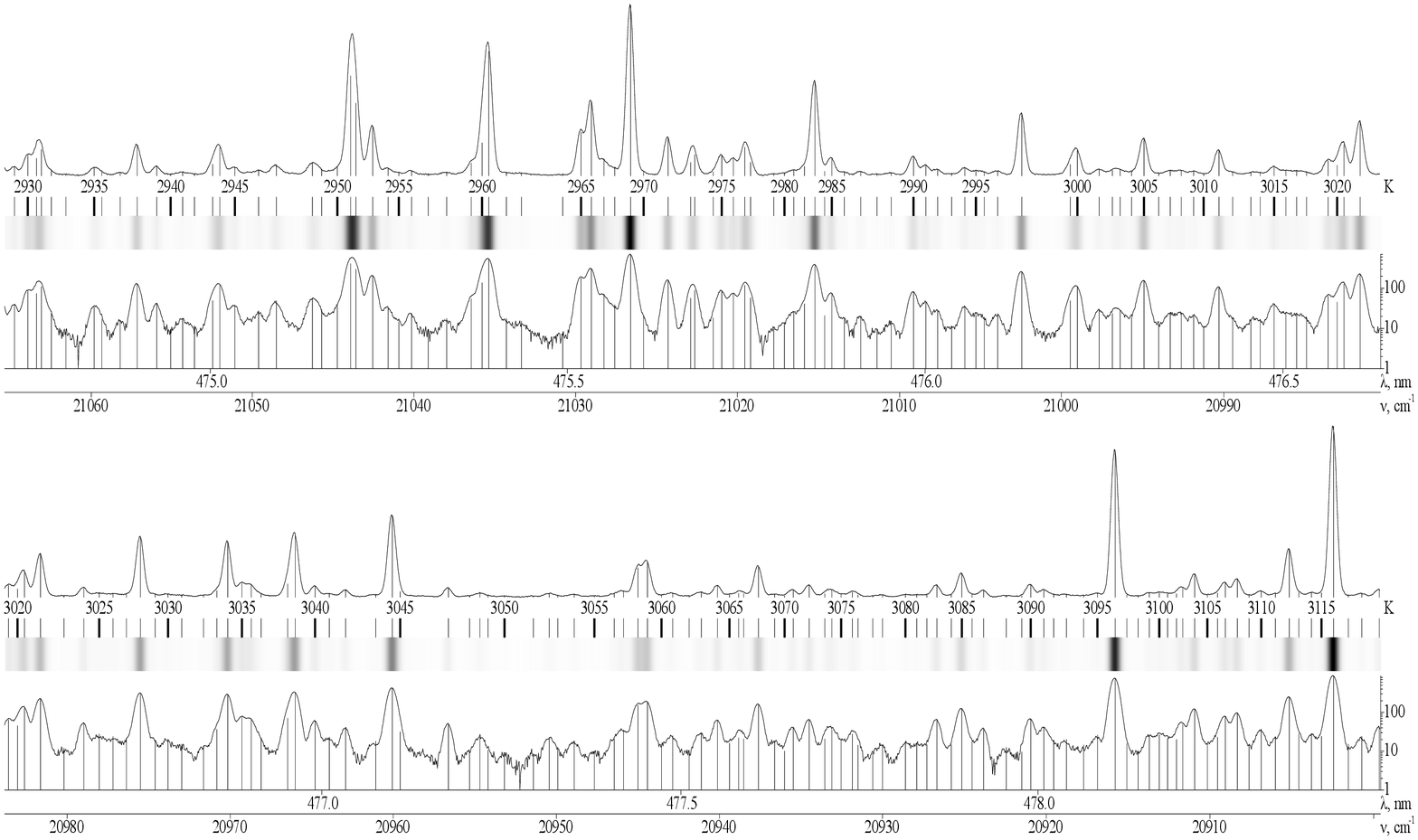}
\end{figure}

\newpage
\begin{figure}[!ht]
\includegraphics[angle=90, totalheight=0.9\textheight]{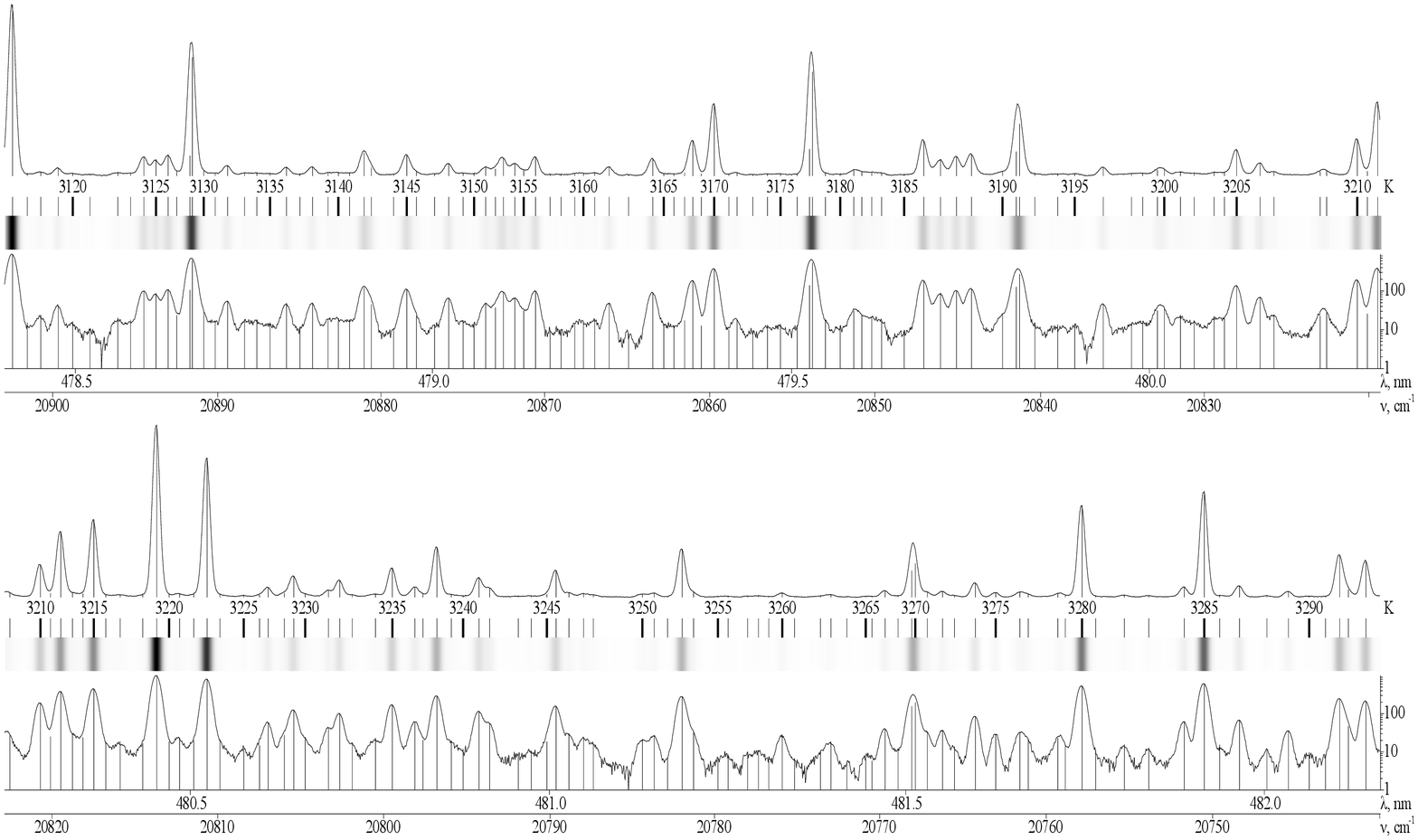}
\end{figure}

\newpage
\begin{figure}[!ht]
\includegraphics[angle=90, totalheight=0.9\textheight]{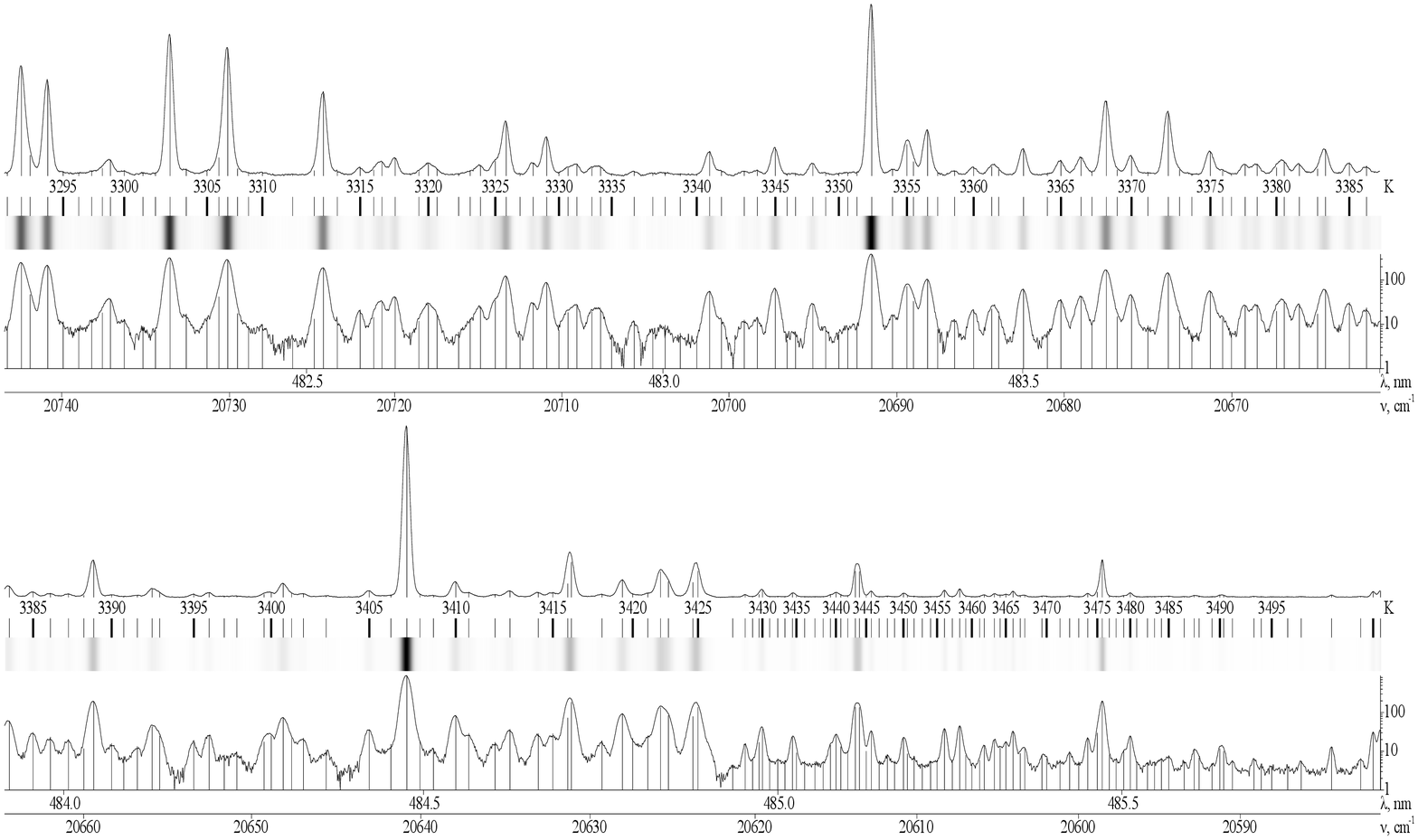}
\end{figure}

\newpage
\begin{figure}[!ht]
\includegraphics[angle=90, totalheight=0.9\textheight]{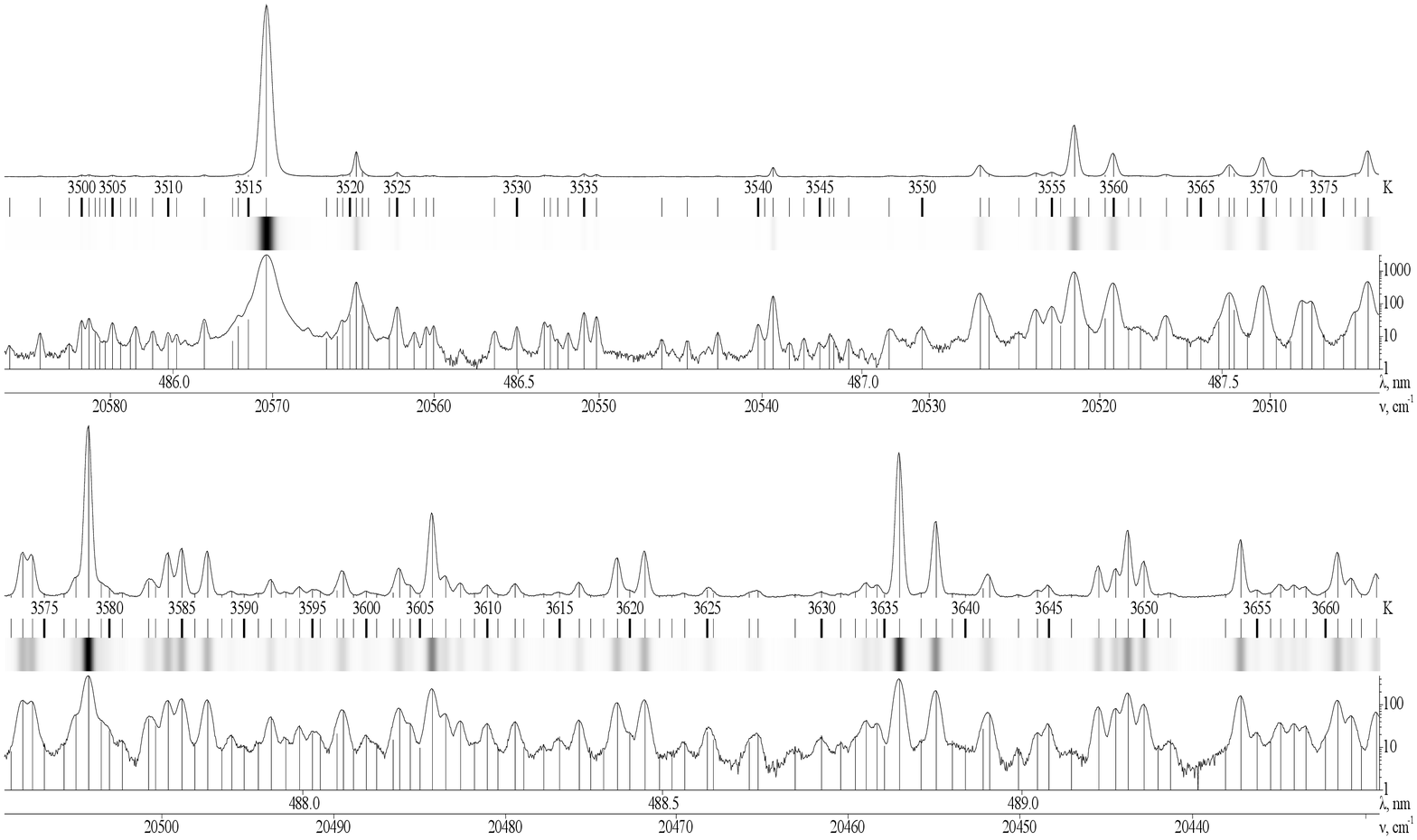}
\end{figure}

\newpage
\begin{figure}[!ht]
\includegraphics[angle=90, totalheight=0.9\textheight]{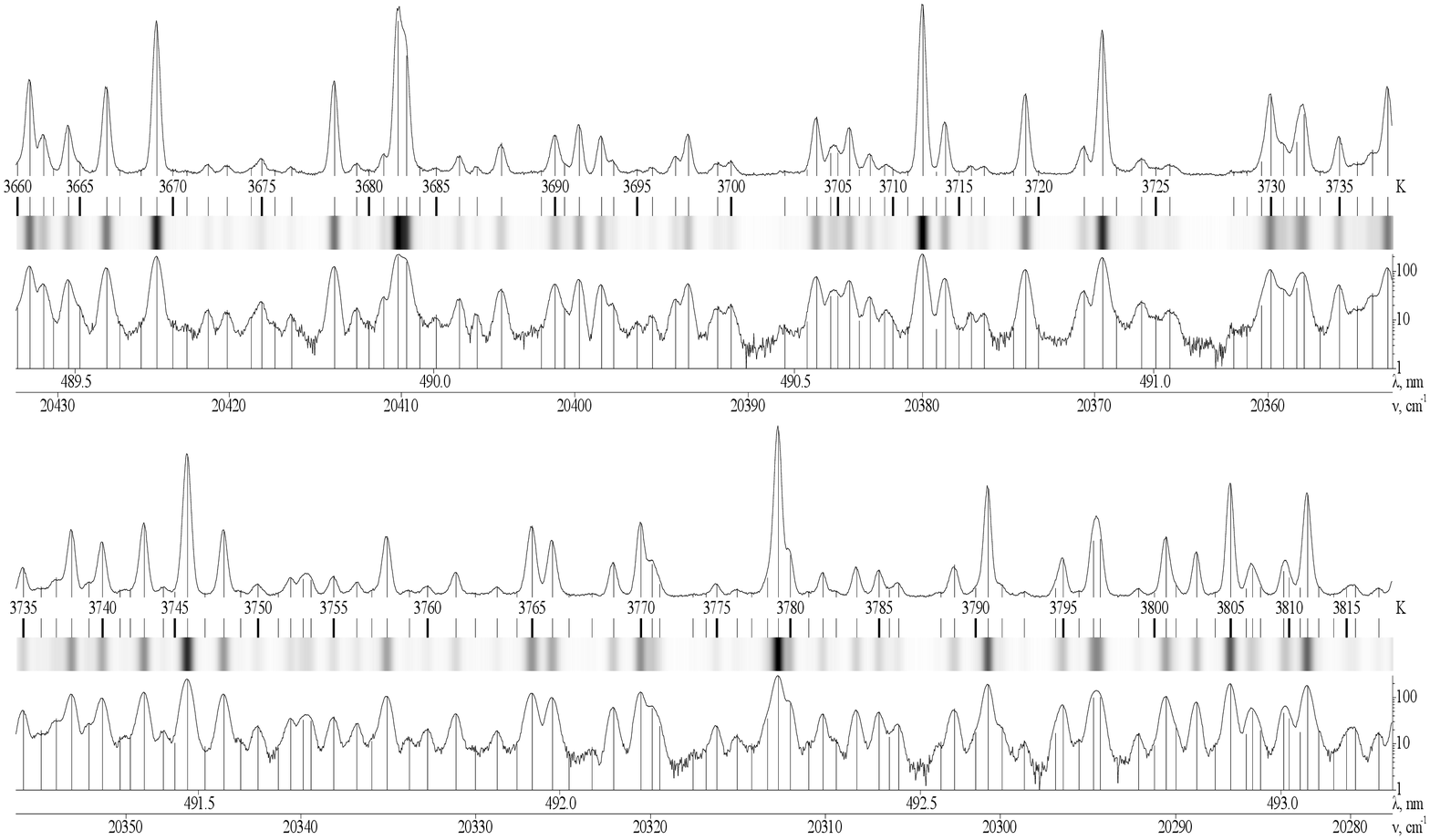}
\end{figure}

\newpage
\begin{figure}[!ht]
\includegraphics[angle=90, totalheight=0.9\textheight]{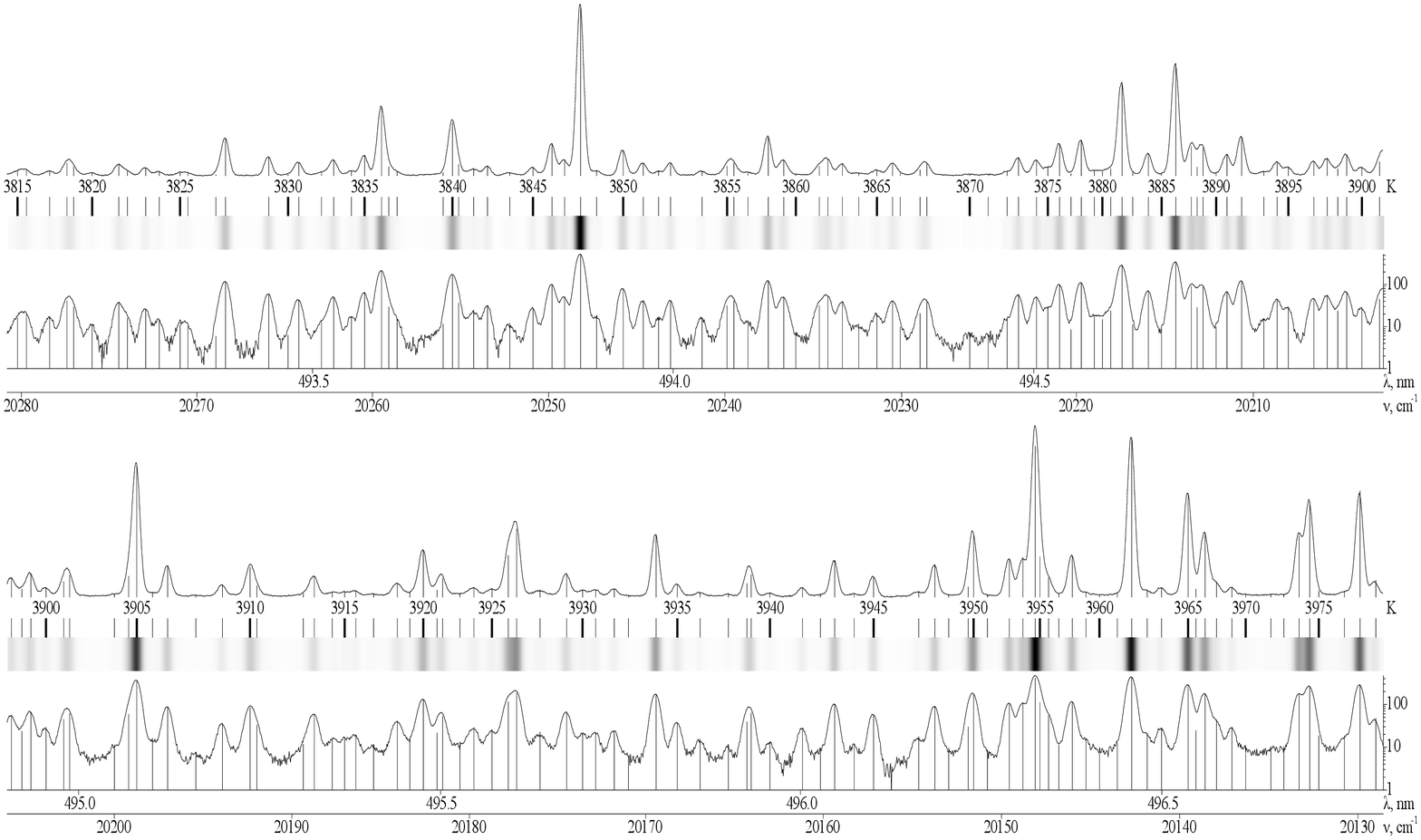}
\end{figure}

\newpage
\begin{figure}[!ht]
\includegraphics[angle=90, totalheight=0.9\textheight]{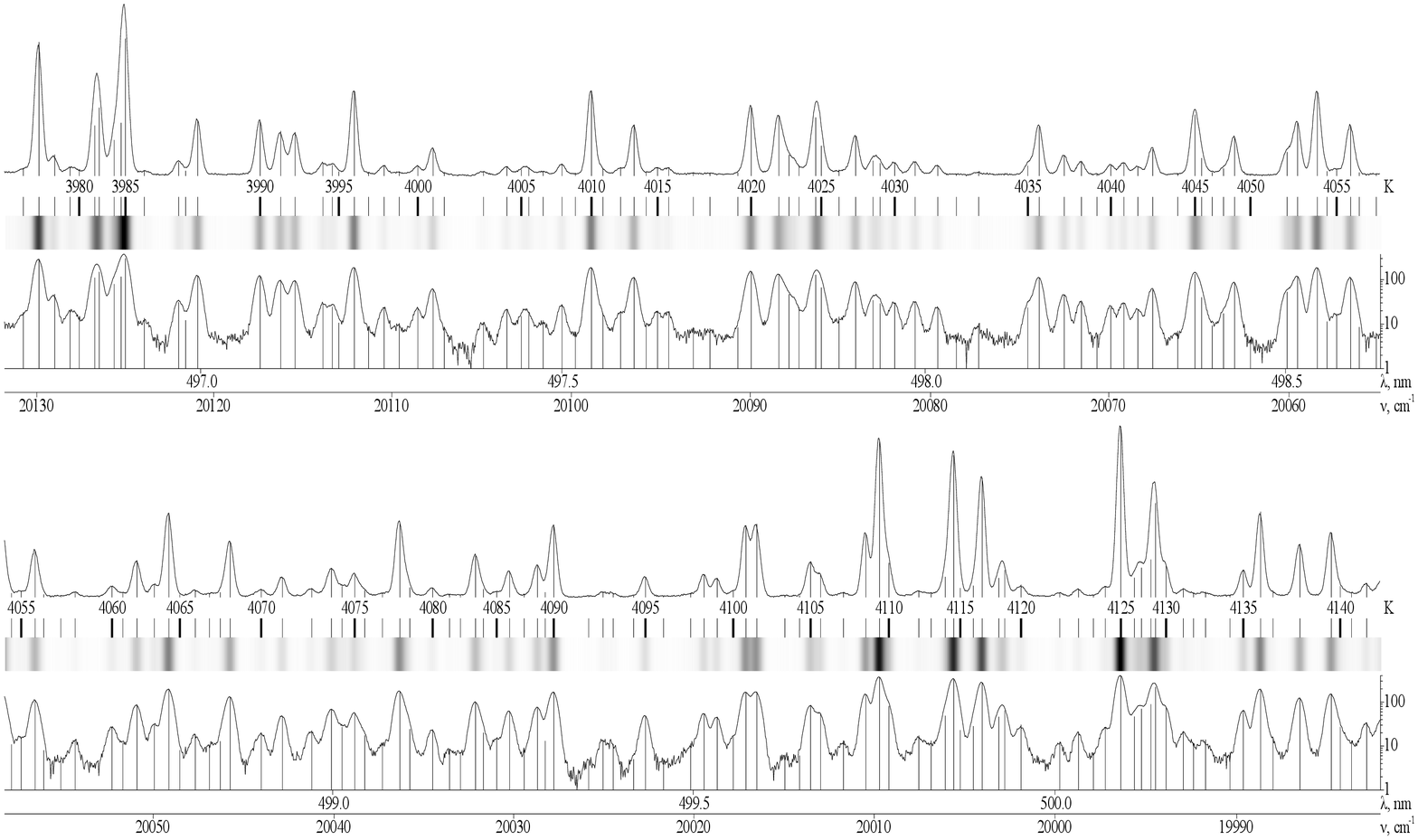}
\end{figure}

\newpage
\begin{figure}[!ht]
\includegraphics[angle=90, totalheight=0.9\textheight]{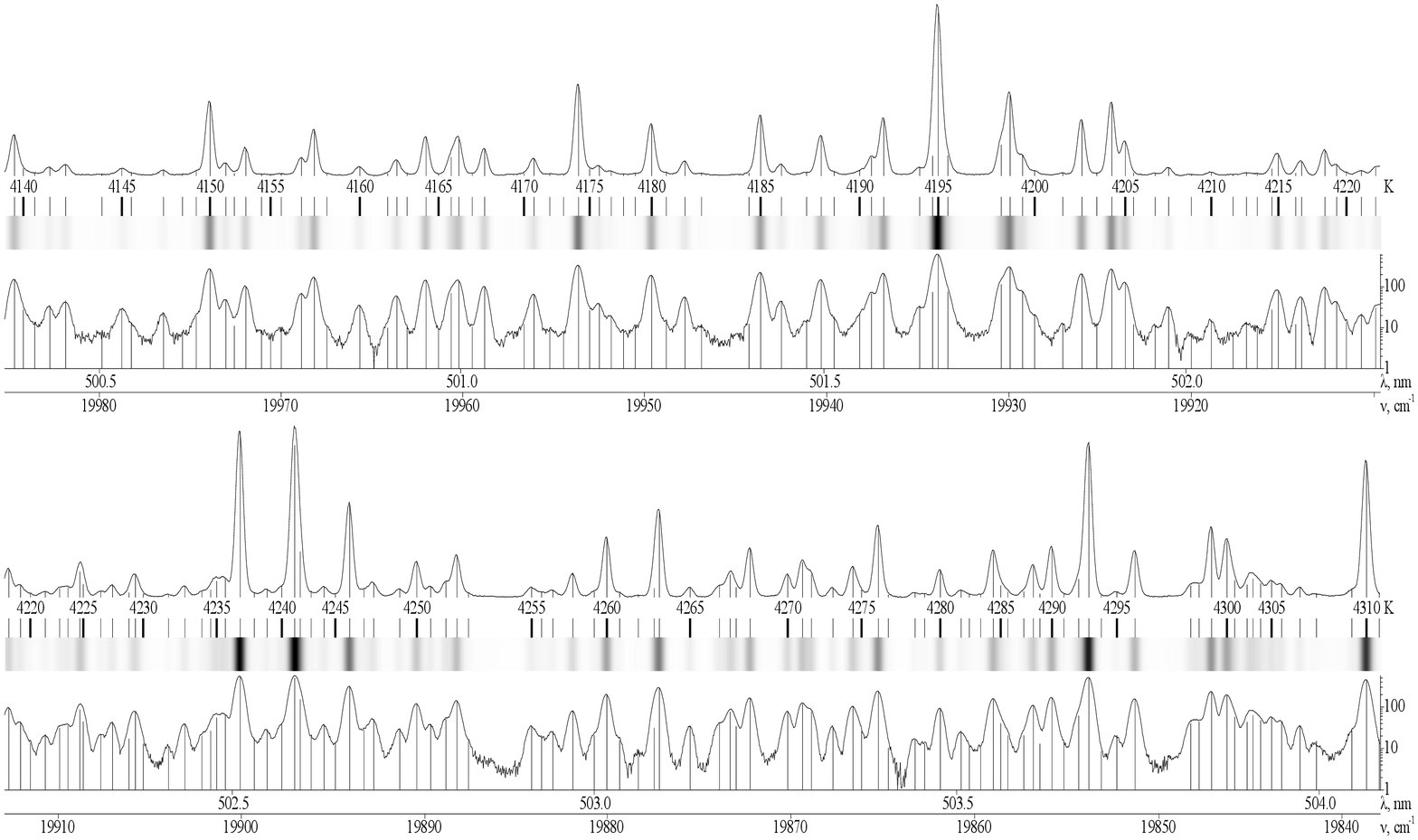}
\end{figure}

\newpage
\begin{figure}[!ht]
\includegraphics[angle=90, totalheight=0.9\textheight]{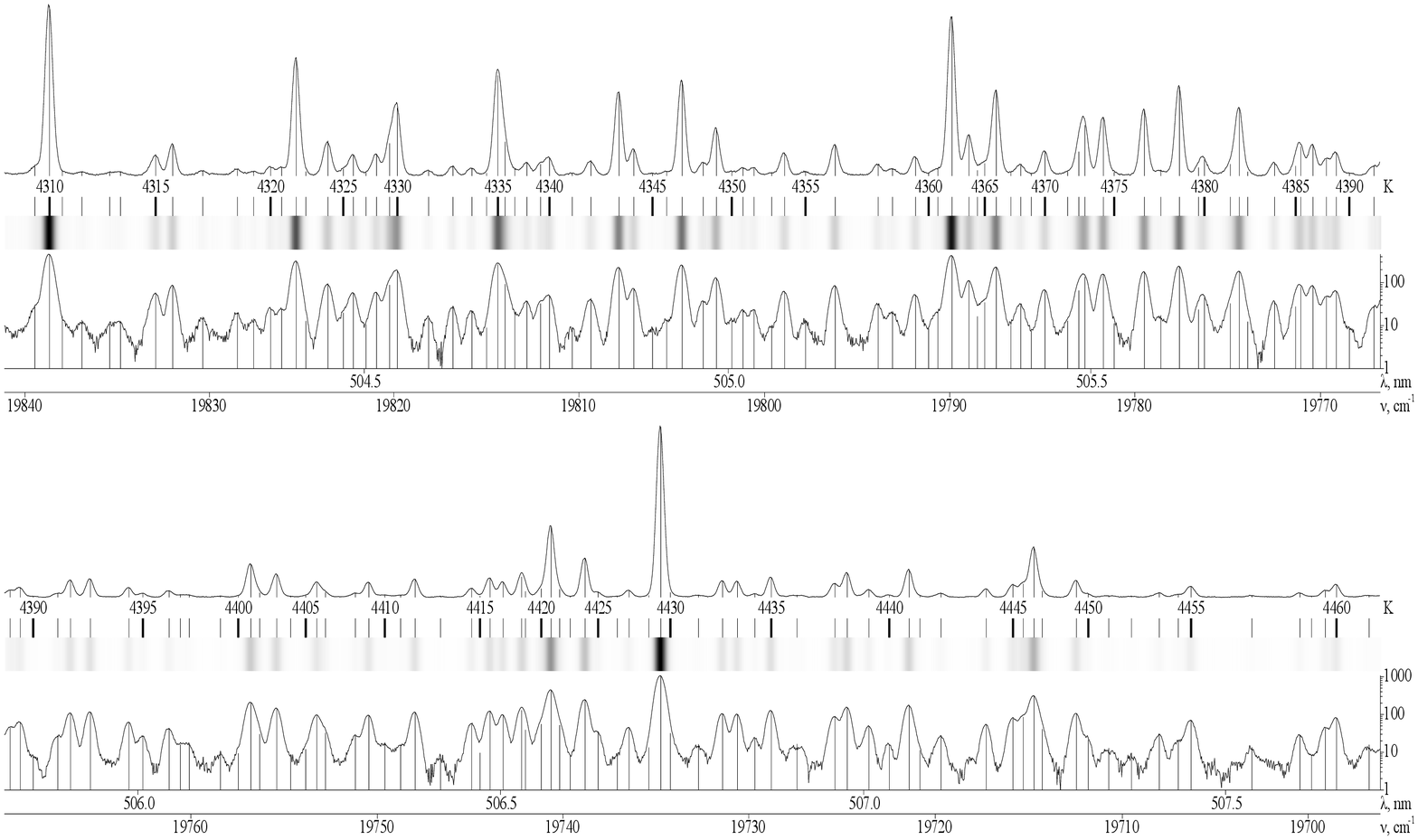}
\end{figure}

\newpage
\begin{figure}[!ht]
\includegraphics[angle=90, totalheight=0.9\textheight]{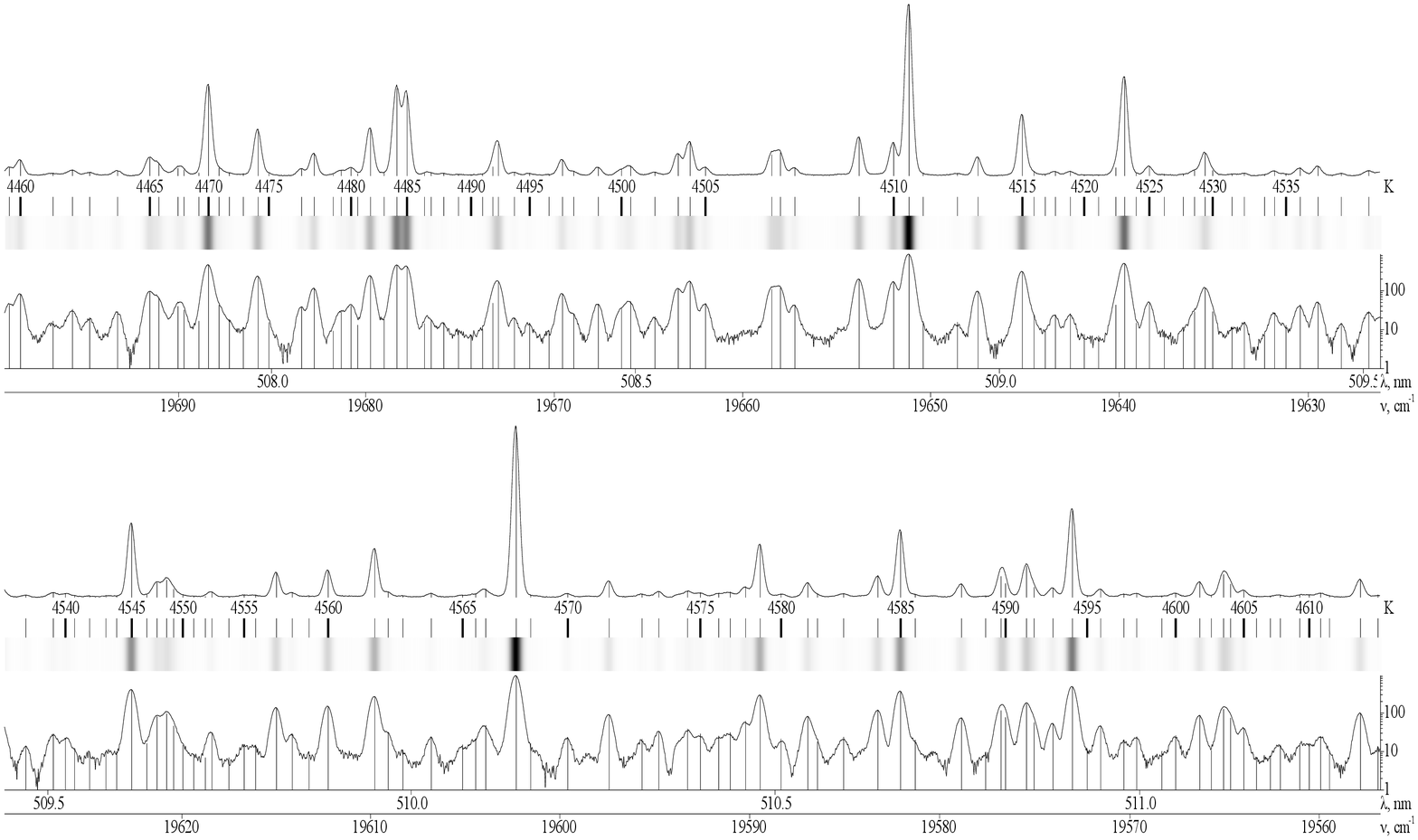}
\end{figure}

\newpage
\begin{figure}[!ht]
\includegraphics[angle=90, totalheight=0.9\textheight]{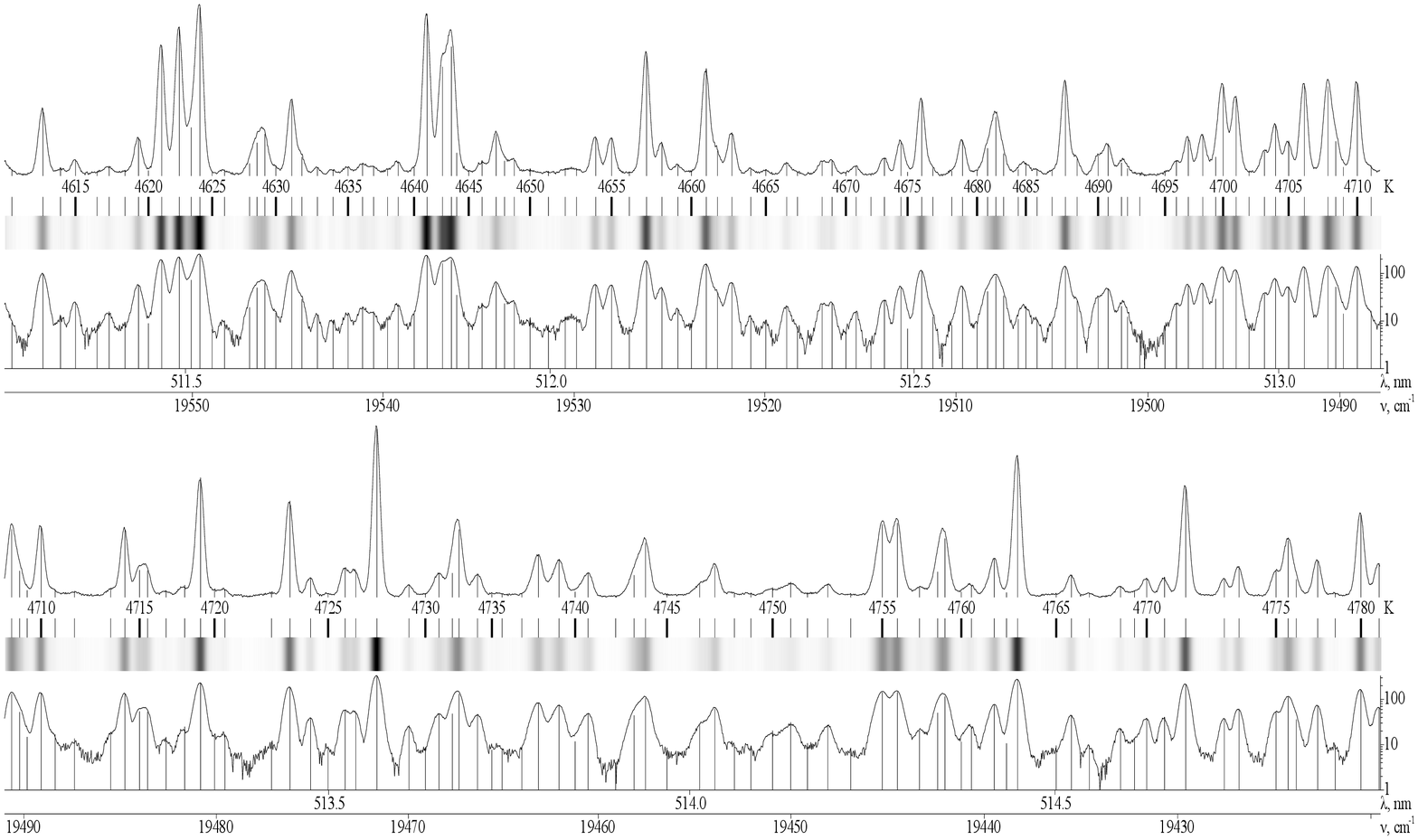}
\end{figure}

\newpage
\begin{figure}[!ht]
\includegraphics[angle=90, totalheight=0.9\textheight]{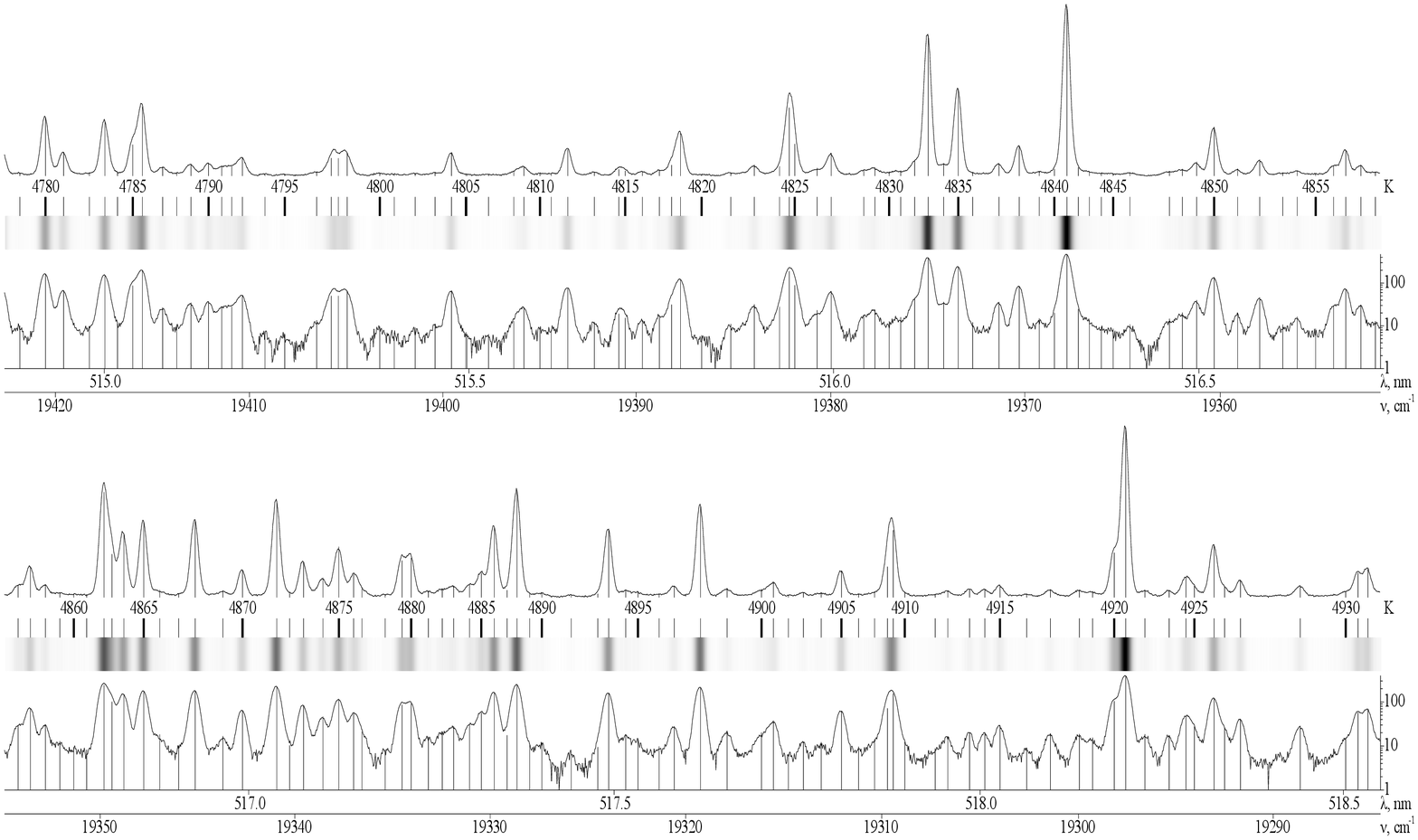}
\end{figure}

\newpage
\begin{figure}[!ht]
\includegraphics[angle=90, totalheight=0.9\textheight]{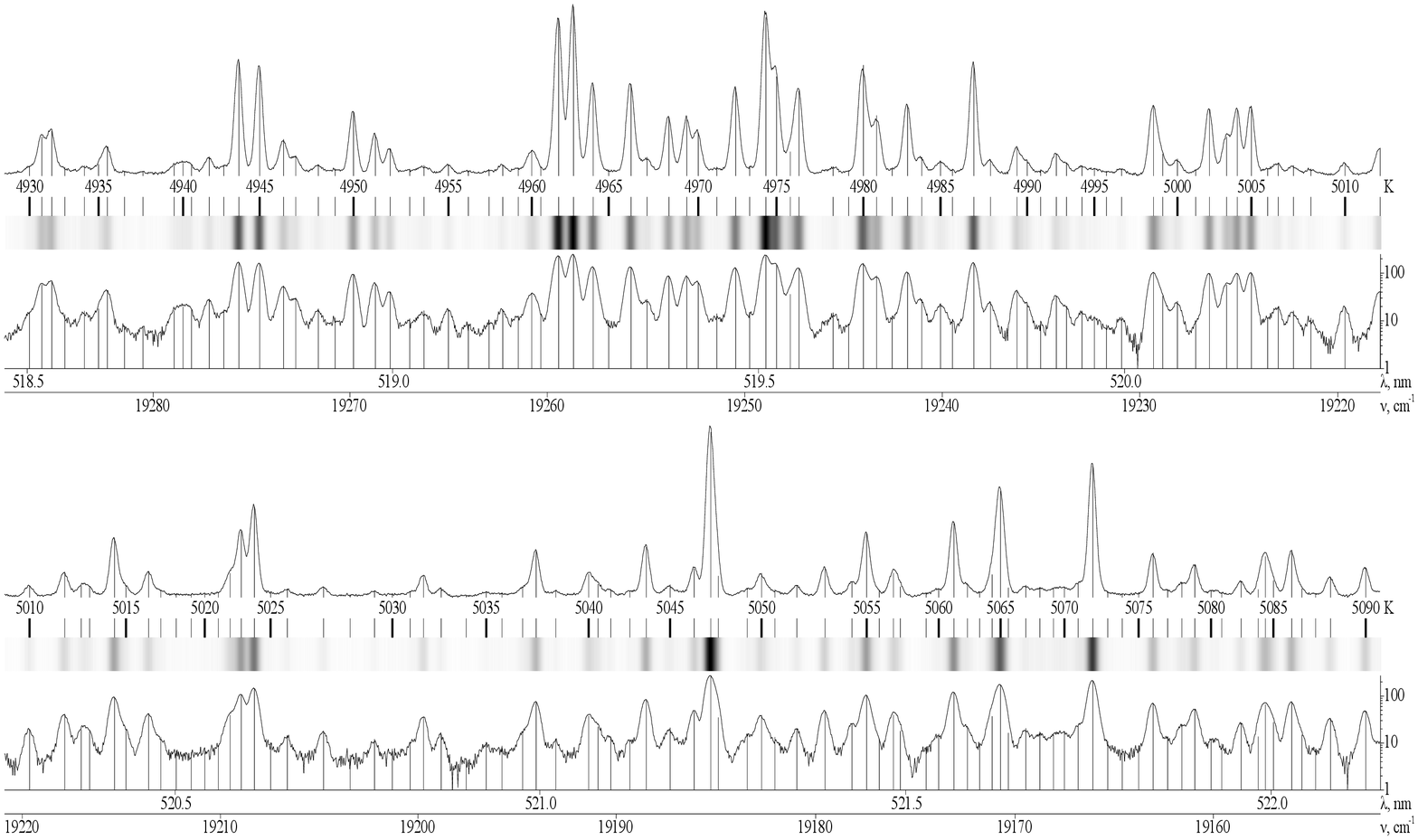}
\end{figure}

\newpage
\begin{figure}[!ht]
\includegraphics[angle=90, totalheight=0.9\textheight]{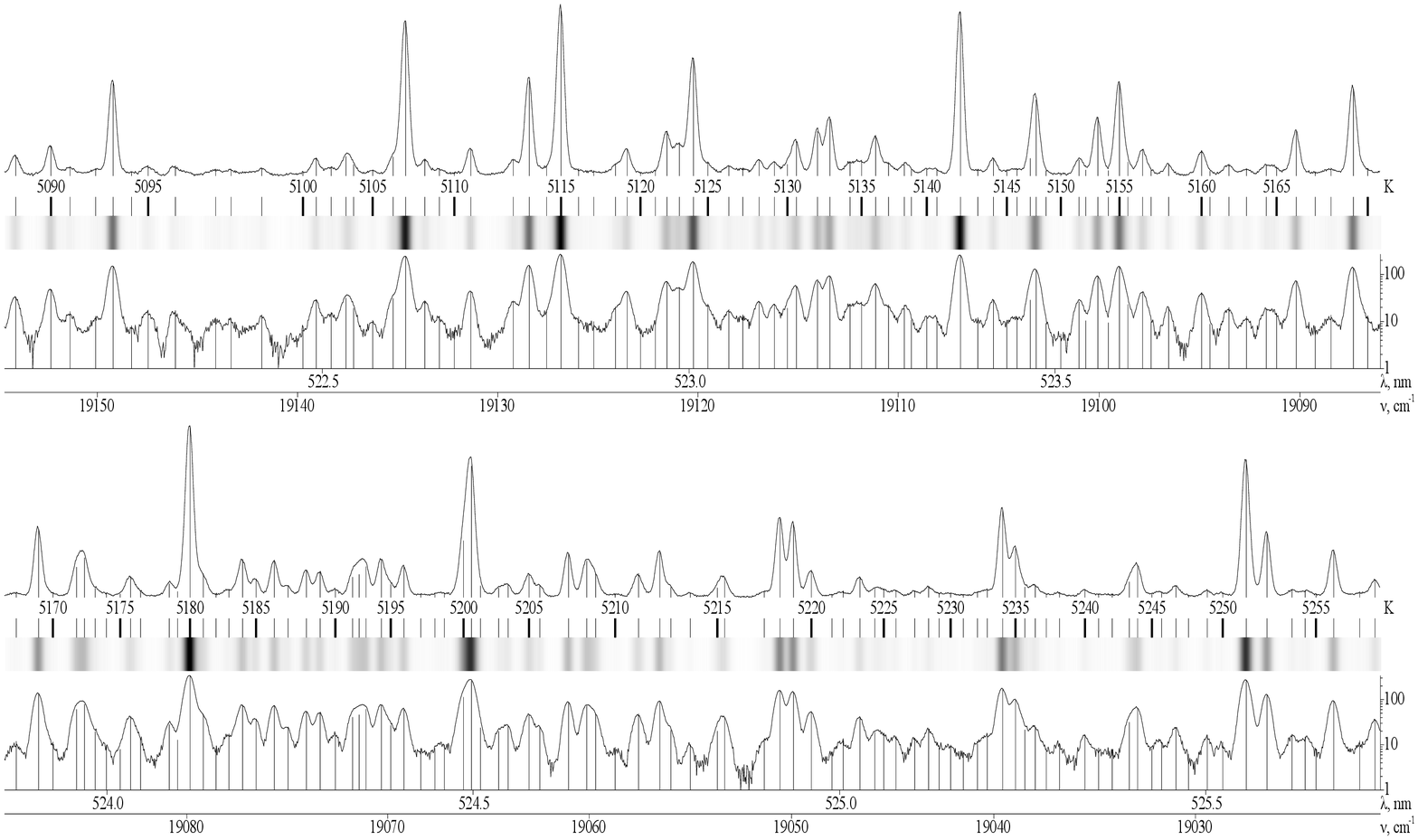}
\end{figure}

\newpage
\begin{figure}[!ht]
\includegraphics[angle=90, totalheight=0.9\textheight]{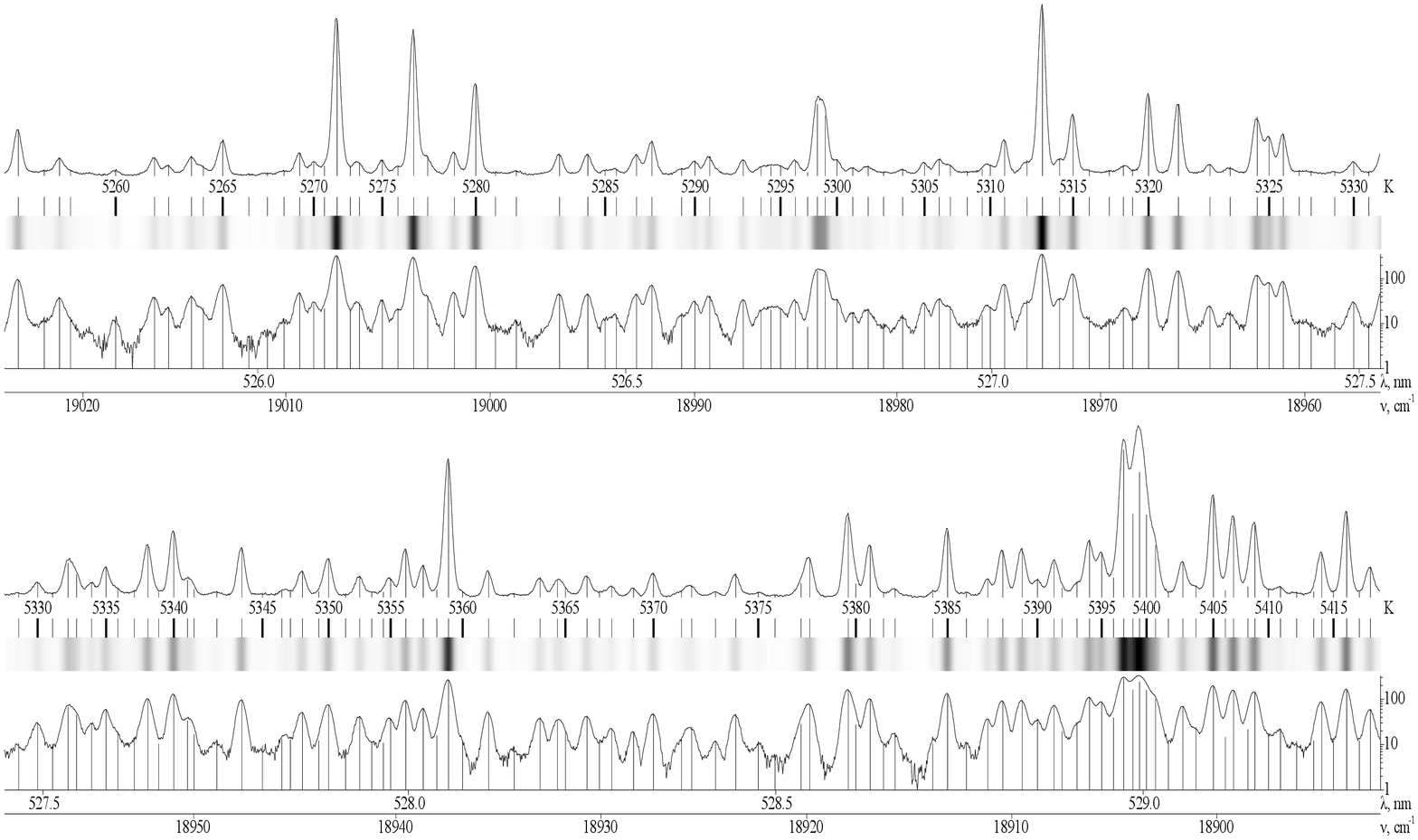}
\end{figure}

\newpage
\begin{figure}[!ht]
\includegraphics[angle=90, totalheight=0.9\textheight]{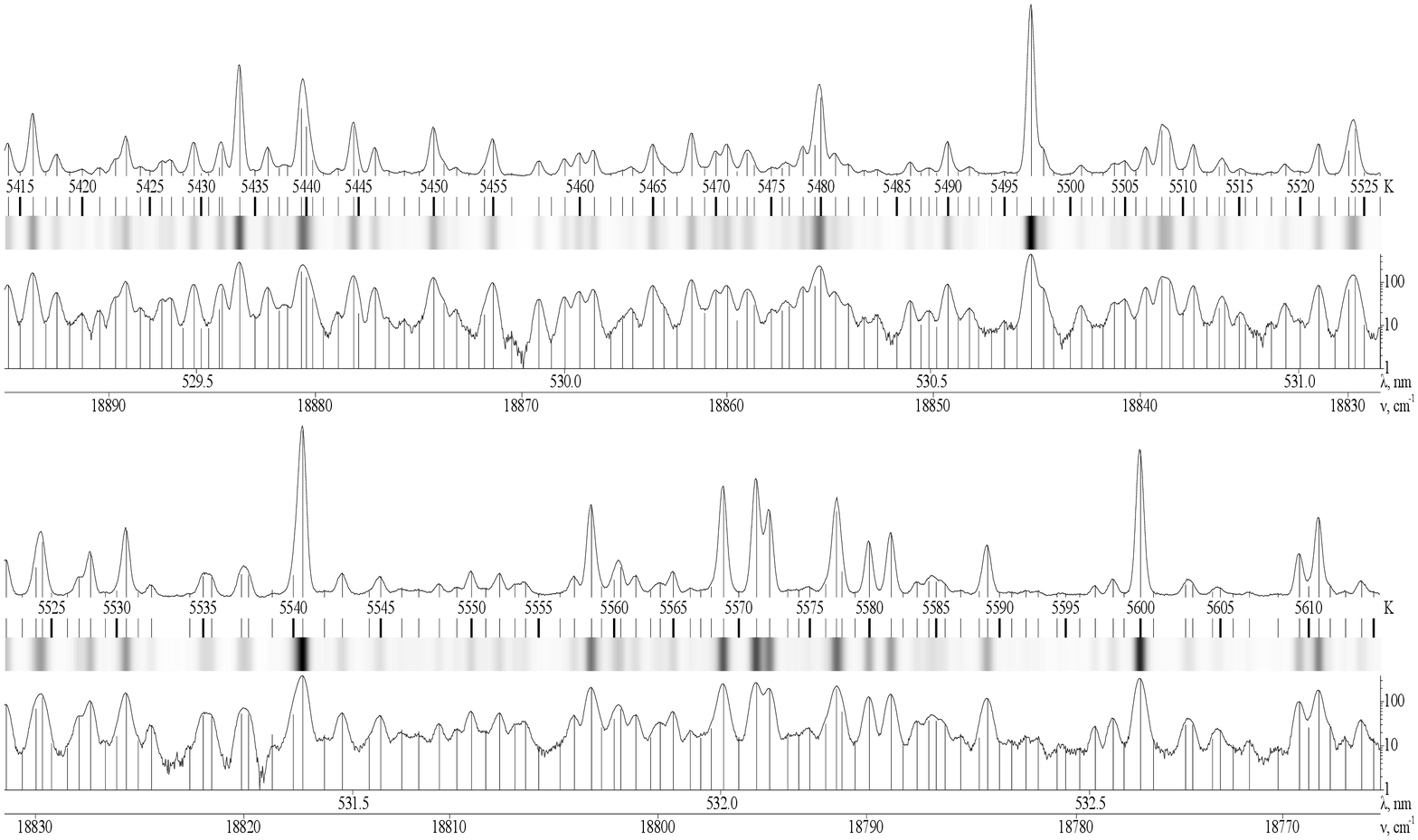}
\end{figure}

\newpage
\begin{figure}[!ht]
\includegraphics[angle=90, totalheight=0.9\textheight]{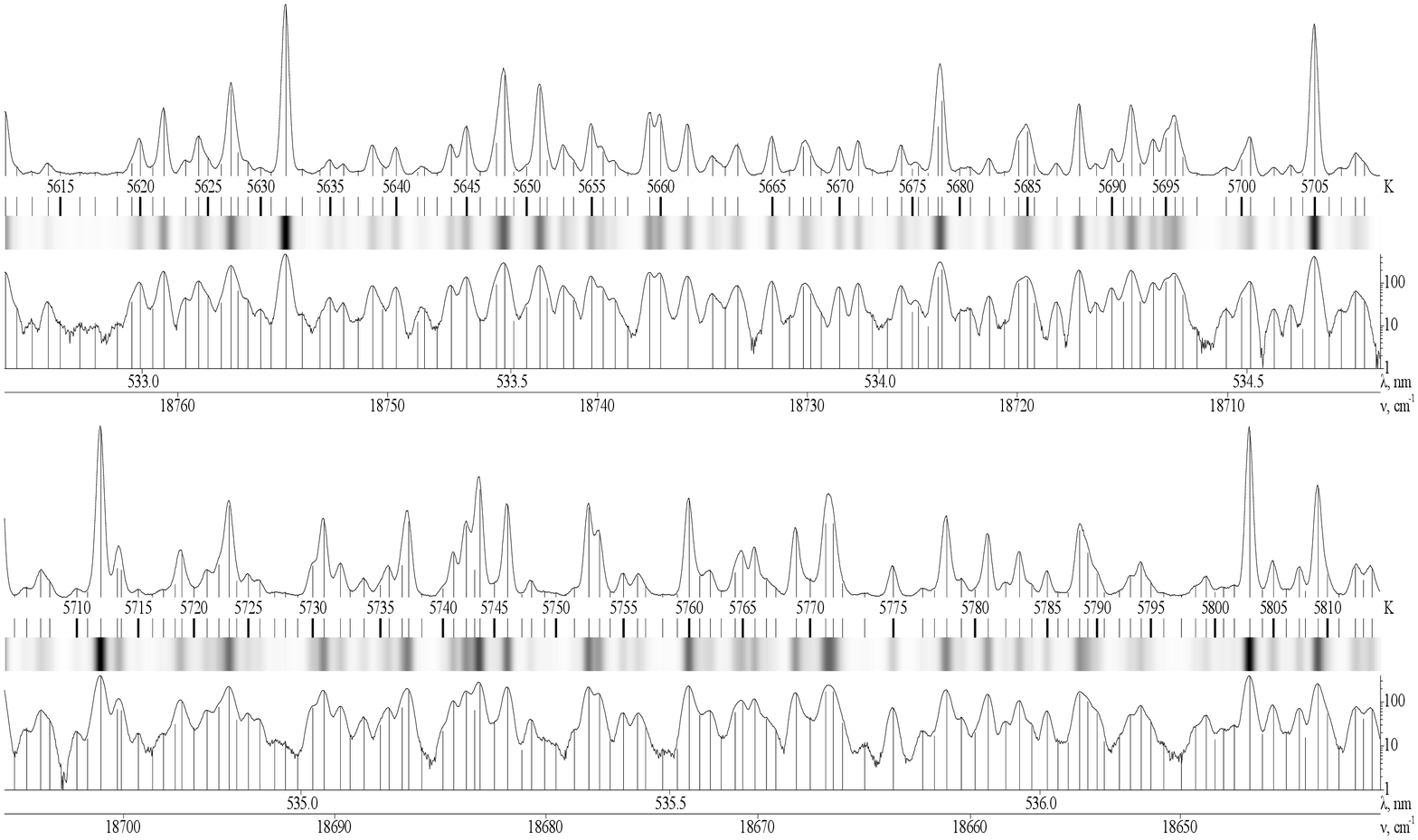}
\end{figure}

\newpage
\begin{figure}[!ht]
\includegraphics[angle=90, totalheight=0.9\textheight]{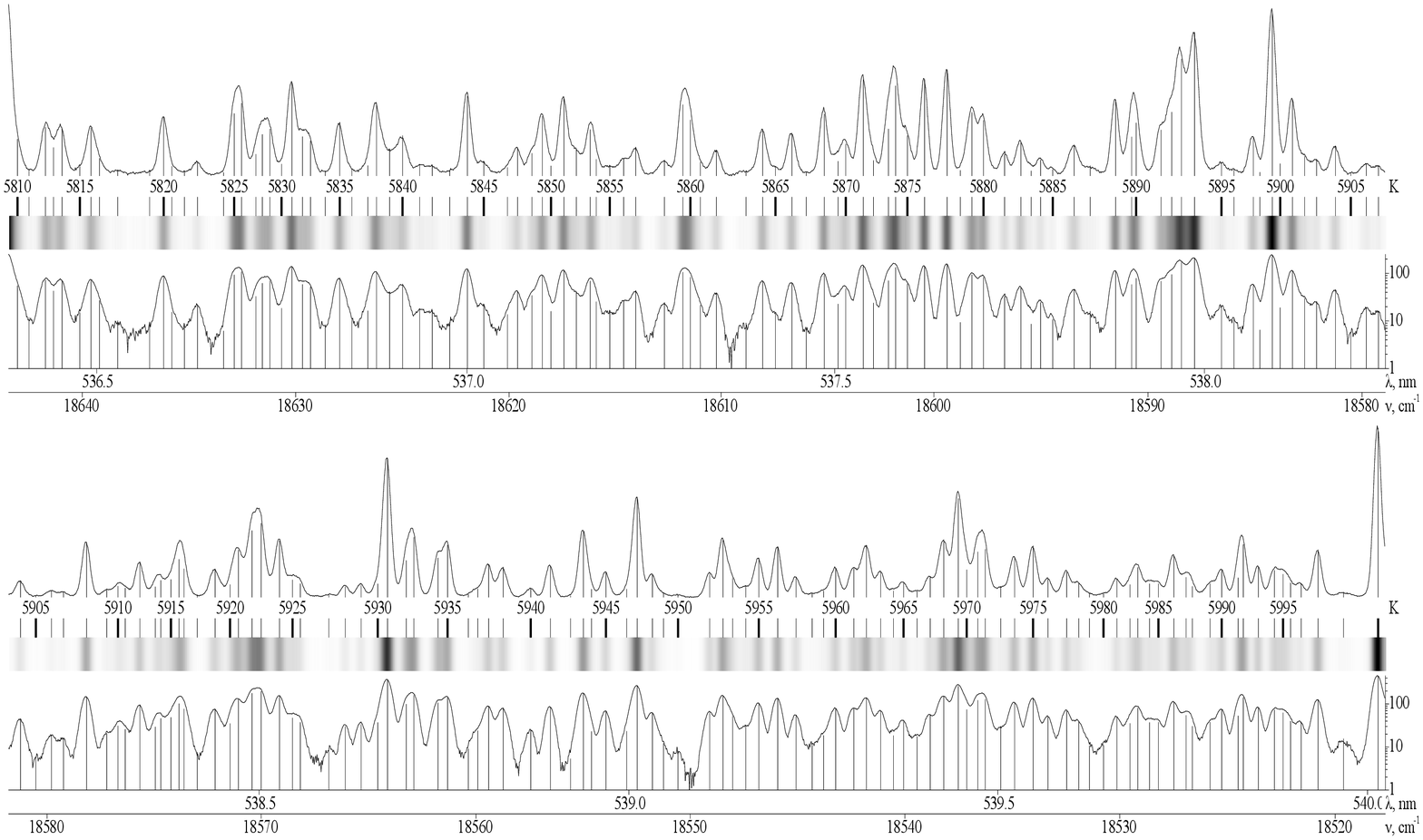}
\end{figure}

\newpage
\begin{figure}[!ht]
\includegraphics[angle=90, totalheight=0.9\textheight]{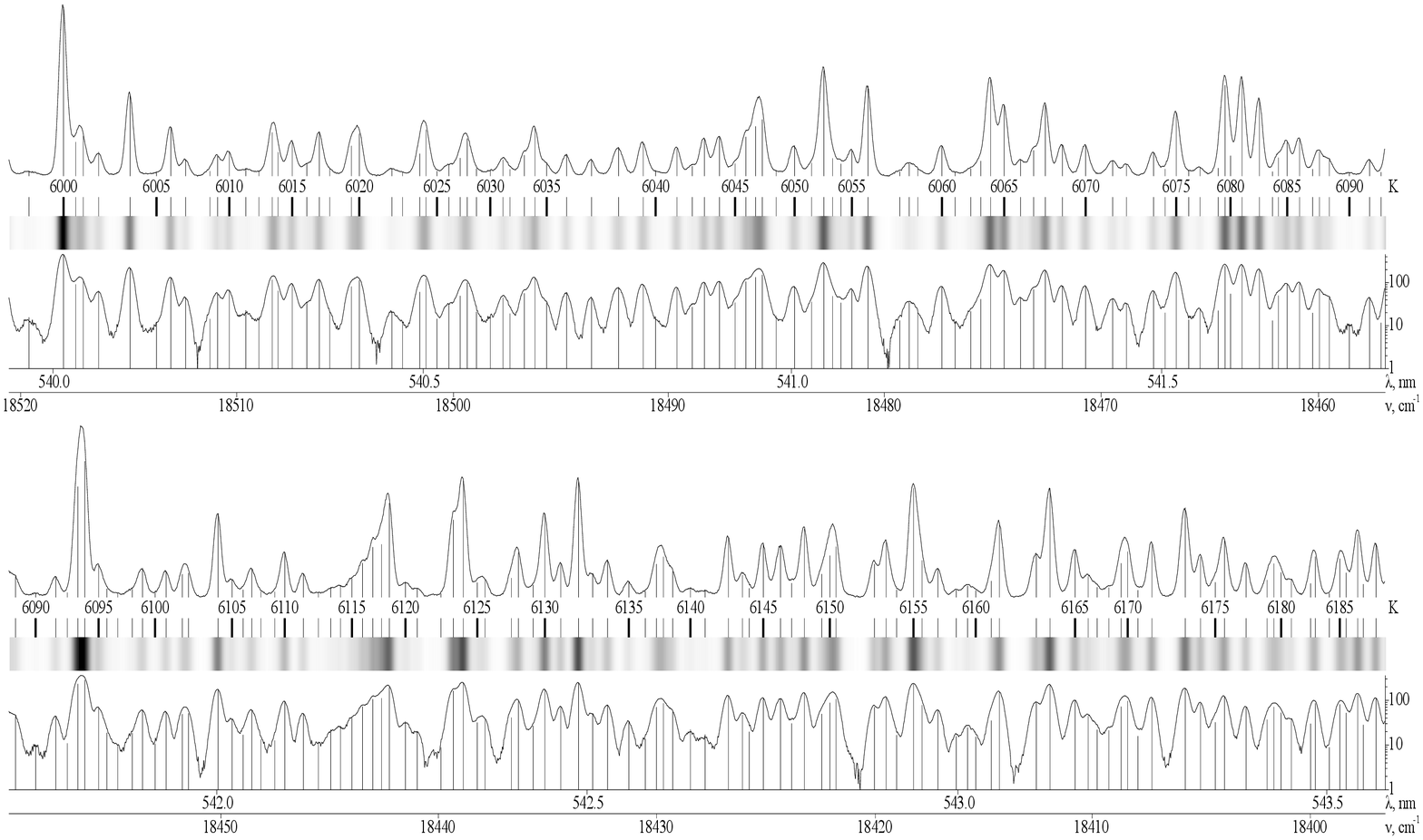}
\end{figure}

\newpage
\begin{figure}[!ht]
\includegraphics[angle=90, totalheight=0.9\textheight]{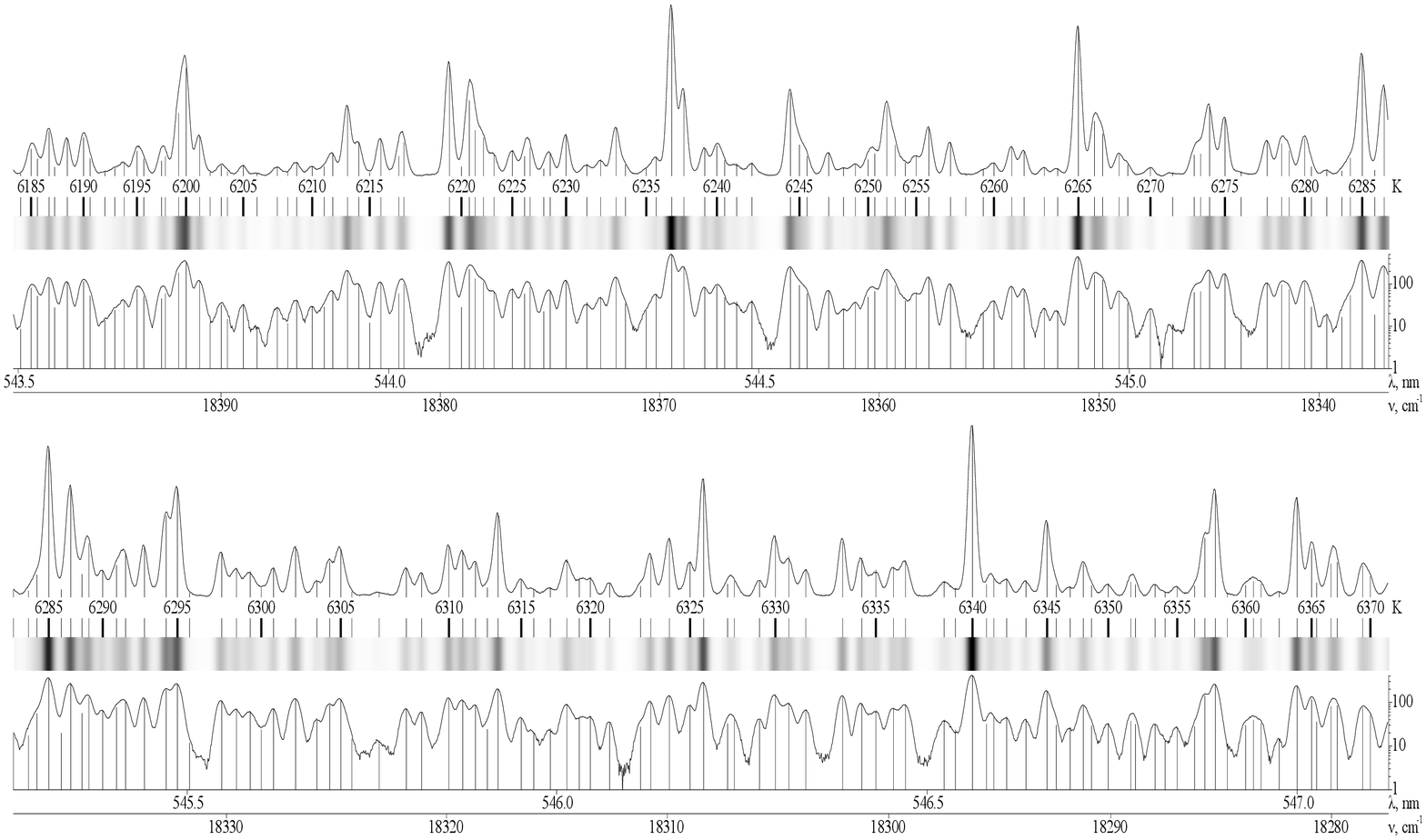}
\end{figure}

\newpage
\begin{figure}[!ht]
\includegraphics[angle=90, totalheight=0.9\textheight]{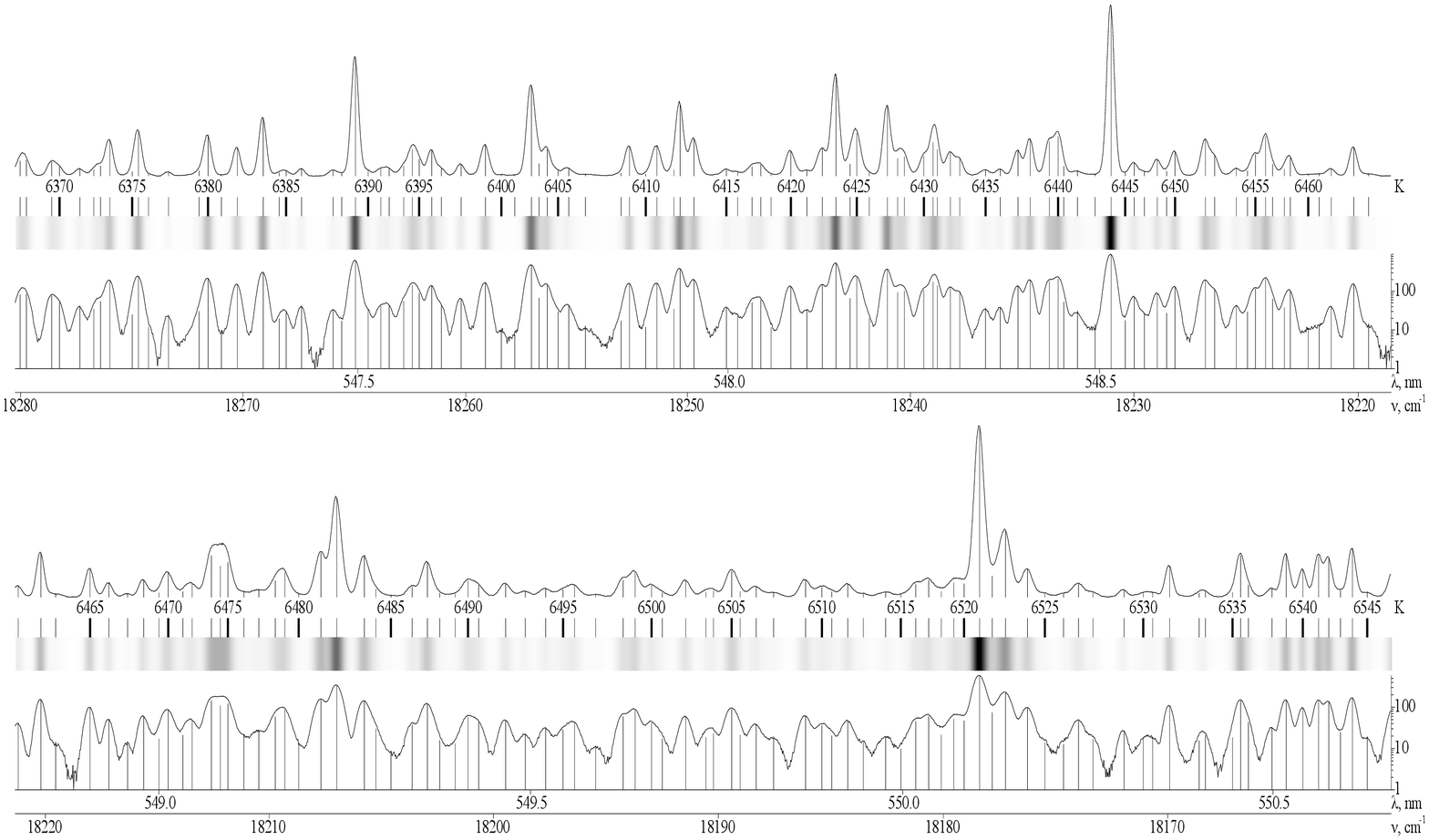}
\end{figure}